\documentclass[aps,reprint,longbibliography]{revtex4-2}
\usepackage{graphicx}
\usepackage{amssymb}
\usepackage{xcolor}
\usepackage[fleqn]{amsmath}
\usepackage[font=small,skip=3pt]{caption}
\usepackage{booktabs}
\PassOptionsToPackage{hyphens}{url}
\usepackage[breaklinks=true]{hyperref}
\usepackage{cleveref}
\usepackage{xurl} 
\usepackage{bm}
\let\mrm\mathrm

\newcommand{\thalf}{t_{1/2}}
\newcommand{\ntn}{\ensuremath{(\mrm{n},2\mrm{n})}}
\newcommand{\np}{\ensuremath{(\mrm{n},\mrm{p})}}
\newcommand{\nna}{\ensuremath{(\mrm{n},\mrm{n}\alpha)}}
\newcommand{\nnp}{\ensuremath{(\mrm{n},\mrm{np})}}

\newcommand{\betam}{\beta^{-}}

\newcommand{\freact}{f_{\text{react}}}



\begin{document}

\title{Production of Nuclear Battery $\beta^{-}$ Emitters Driven by Fusion Neutrons}

\author{J.~F.~Parisi}
\email{jason@marathonfusion.com}
\affiliation{Marathon Fusion, 150 Mississippi Street, San Francisco, CA 94107, USA}

\begin{abstract}
Nuclear batteries require radioisotopes with specific combinations of half-life, decay mode, and radiation properties, yet most candidate fuels lack scalable production routes. We show how the future availability of deuterium-tritium (D-T) fusion neutrons could enable manufacturing nuclear battery radioisotopes at many orders of magnitude higher rate than at present. We assess the capability of 14 MeV D-T fusion neutrons to produce nuclear battery radioisotopes by simulating feedstock material irradiation with neutrons. Promising radioisotope candidates include ${}^{147}$Pm, ${}^{63}$Ni, ${}^{39}$Ar, and ${}^{137}$Cs. Some feedstocks allow a radioisotope to be produced at scale while also closing the tritium fuel cycle, resulting in hundreds to over one thousand kilograms of high specific activity radioisotope per gigawatt thermal year of D-T fusion irradiation. We perform OpenMC simulations of an enriched ${}^{148}$Nd blanket for a tokamak, demonstrating that tritium self-sufficient designs can produce over one ton of ${}^{147}$Pm per gigawatt thermal year, equivalent to $\sim$one billion Curies per year of ${}^{147}$Pm. Operation of such a blanket would represent an unprecedented increase of nuclear battery radioisotope production capability.
\end{abstract}

\maketitle

\section{Introduction}

Nuclear batteries convert radioactive decay energy into electricity. Radioisotopes store energy at densities up to $\sim 2 \times 10^{9}$~J\,g$^{-1}$ (${}^{238}$Pu), roughly five million times that of lithium-ion batteries~\cite{Prelas2016}. Several classes of nuclear batteries exist~\cite{Prelas2016,Olsen2012}, including radioisotope thermoelectric generators (RTGs), which convert decay heat to electricity through thermocouples~\cite{OBrien2008,Dustin2021}; betavoltaic and alphavoltaic cells, which convert charged-particle energy directly in a semiconductor junction~\cite{Olsen2012,Krasnov2021}; and thermophotovoltaic~\cite{Lapotin2022} and direct-charge-collection~\cite{Wang2019} devices. Each architecture places different demands on the fuel isotope: RTGs favor alpha emitters with high specific thermal power $P_{\text{th}}$, whose short-range particles self-absorb in the fuel form and simplify containment~\cite{OBrien2008,Rinehart2001}, whereas betavoltaic and direct-charge devices require low-energy beta emitters ($E_{\beta,\max} \lesssim 300$~keV) with negligible co-emitted gamma radiation to minimize semiconductor displacement damage~\cite{Olsen2012,Krasnov2021}. This work is concerned with the production of suitable beta-emitting radioisotopes rather than with any particular battery architecture.

The supply of suitable isotopes is constrained. The dominant RTG fuel, ${}^{238}$Pu, is produced at ${\sim}\,1.5$~kg\,yr$^{-1}$ via ${}^{237}$Np irradiation at the High Flux Isotope Reactor ~\cite{Cataldo2020,SpaceNewsPu238,ORNLPu238,INLPu238}. The European Space Agency is pursuing ${}^{241}$Am extracted from aged PuO$_2$ stockpiles as an alternative RTG fuel~\cite{Ambrosi2019,AMPPEX}, with production capacity of ${\sim}\,500$~g\,yr$^{-1}$. Betavoltaic fuels face even greater constraints: ${}^{63}$Ni production requires irradiation of enriched ${}^{62}$Ni in high-flux reactors followed by isotopic enrichment, with only a handful of facilities worldwide capable of making it~\cite{Ni63Production}; current US inventories are at the Curie level, far short of the tens of Curies per unit that a single commercial 1~W betavoltaic device would require~\cite{IEEESpectrum2025}. ${}^{14}$C diamond betavoltaic prototypes have been demonstrated at picowatt scale~\cite{Arkenlight2024}, and tritium-based betavoltaics are commercially available but limited in power density~\cite{CityLabs}. Expanding domestic radioisotope production infrastructure is a priority for the U.S.\ Department of Energy, with the planned Radioisotope Processing Facility at Oak Ridge National Lab~\cite{DOEIsotope2025,ANSIsotopeSupply}. While device-level challenges remain (conversion efficiency, radiation damage, regulatory frameworks), a fundamental barrier to scaling is radioisotope supply~\cite{Prelas2016,IEEESpectrum2025,ChemWorld2024}. The emerging fusion neutron industry, with its unique access to fast neutron-driven reactions could provide the supply-side breakthrough that opens an expanded commercialization path for nuclear batteries. In this article we focus on $\betam$ emitters; $\alpha$ emitters will be covered in future work.

\begin{figure*}[htbp]
\centering
\includegraphics[width=2\columnwidth]{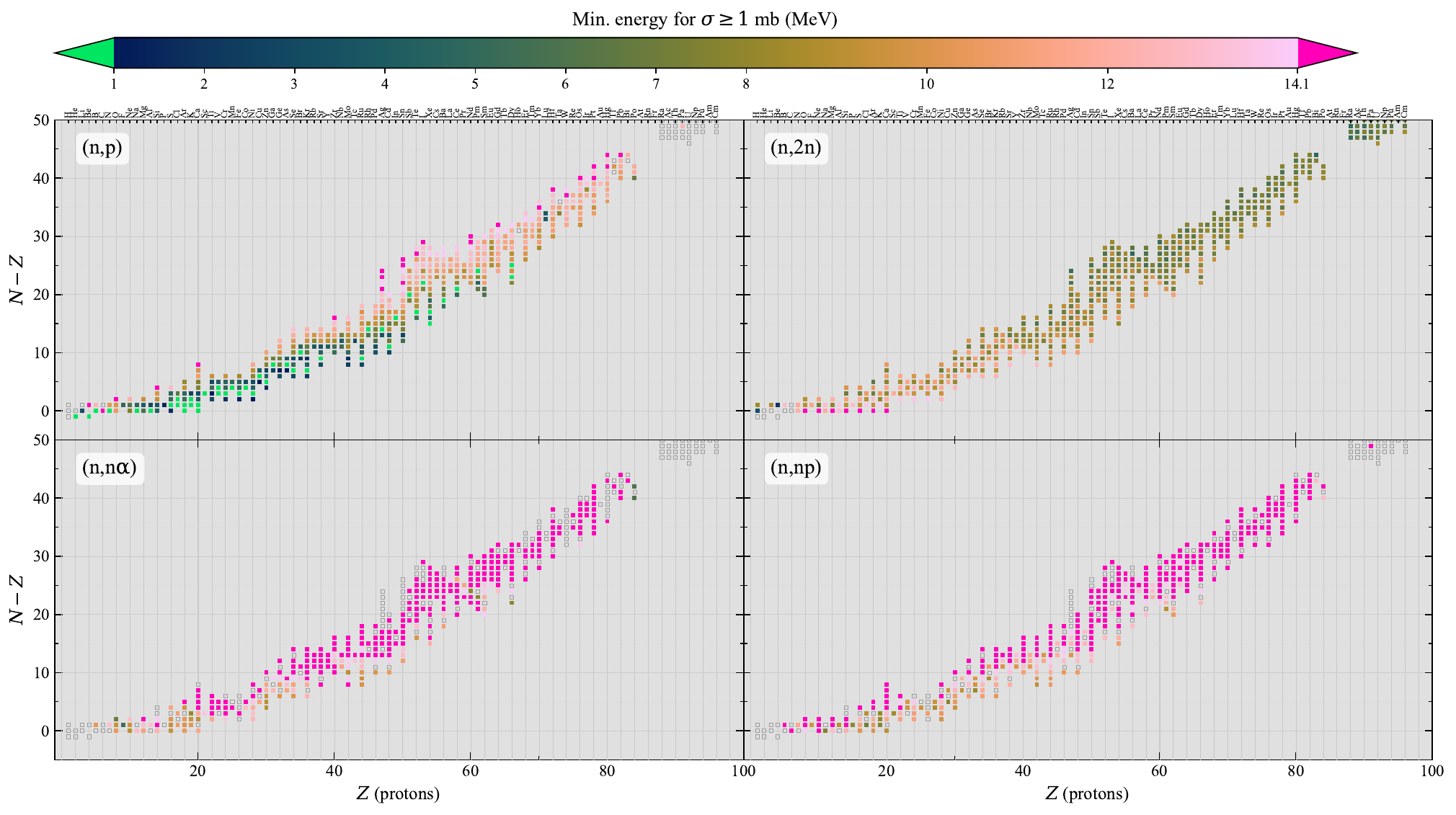}
\caption{Required neutron energy for cross section to exceed 1 millibarn on a target isotope. A blank spot means the reaction never exceeds 1 millibarn.}
\label{fig:threshold_reactions}
\end{figure*}

Fast fusion neutrons are uniquely suited for producing many nuclear battery radioisotopes at scale. This follows from two key properties of fast fusion neutrons: (i) a large fraction of source neutrons, often as high as $\sim$50-60\% can drive the desired transmutation pathway, and (ii) the large number of free neutrons \textit{available} for transmutation. Property (i) differentiates fusion neutron-driven transmutation from charged particle-driven transmutation because neutrons do not experience the Coulomb force when damping in materials, whereas the overwhelming majority of charged-particles will be lost due to Coulomb collisions. Property (ii) differentiates fusion neutrons from both charged-particle and fission-neutron driven transmutation: charged particle sources seldom exceed 1 milliamp ($6 \cdot 10^{15}$ protons/s) whereas a single megawatt of deuterium-tritium (D-T) fusion power produces $3.6 \cdot 10^{17}$ neutrons/s, and in fission systems there is a highly engineered, precise neutron economy whereupon an extensive re-design would be required to allow such systems to dedicate a large fraction of fission-born neutrons to drive transmutation using a large fraction of neutrons produced in fission reactions. In fusion systems producing isotopes with (n,2n) reactions, all of the fast neutrons can be dedicated to isotope production while still satisfying other system requirements such as tritium self-sufficiency~\cite{Rutkowski2025}. In smaller ($\lesssim$MW-class) fusion systems, tritium self-sufficiency is less important because one megawatt year of fusion power burns only 56 grams of tritium, a quantity that is relatively straightforward to source externally and is justified economically if the radioisotope is much more valuable than tritium. 

D-T fusion devices produce 14.1~MeV neutrons that access reactions such as $\ntn$, $\np$, $\nnp$, and $\nna$ ~\cite{Gilbert2024,Parisi2025,Evitts2025,Rutkowski2025}. These reactions are typically energetically forbidden or have low cross section at neutron energy $\lesssim$1 MeV. In \Cref{fig:threshold_reactions} we show the minimum neutron energy required for a reaction cross section greater than 1 millibarn - most of these reactions require neutrons with at least several MeV of energy. Given that most of these cross sections are vanishingly small for slower neutrons, D-T neutrons often have a significant advantage over irradiation by fission neutrons for reaction pathways that change the proton number. Furthermore, because most reaction chains we consider here change the proton number, the product can be chemically separated from the feedstock, producing high specific activity that nuclear batteries require. Fusion may therefore be the key to breaking the radioisotope supply bottleneck that currently confines nuclear batteries to niche applications, especially for betavoltaics.

This article is structured as follows. In \Cref{sec:landscale} we introduce nuclear battery radioisotope landscape. We calculate yields on feedstocks irradiated by fusion neutrons in \Cref{sec:fusion_yields}. \Cref{sec:scale} shows how the availability of nuclear battery radioisotope can scale with the fusion industry. We discuss compatibility with tritium self-sufficiency in \Cref{sec:neutron}. Production pathways with lower specific activity are briefly discussed in \Cref{sec:deltaz0_n2n}. The value per fusion neutron of producing nuclear battery radioisotopes is calculated in \Cref{sec:vpn}. We summarize in \Cref{sec:discussion}. Depletion neutronics simulations are described in Appendix \ref{app:tokamak_PmSm} and some neutron cross sections are given in Appendix \ref{app:more_cross_sections}.

\section{Nuclear battery radioisotope landscape} \label{sec:landscale}

A radioisotope with decay constant $\lambda = \ln 2/ \thalf$ and half-life $\thalf$ has specific power
\begin{equation}
P_\mathrm{th} (t) = \lambda e^{- \lambda t} E_\mathrm{decay} \frac{N_\mathrm{A} }{M_\mathrm{mol}},
\end{equation}
where $t$ is time after production, $E_\mathrm{decay}$ is the deposited energy per decay, and $M_\mathrm{mol}$ is the molar mass. Crucially, throughout this work $E_\mathrm{decay}$ denotes the total decay energy $Q$ minus the neutrino energy $E_\mathrm{\nu}$ and photon energy $E_\mathrm{\gamma}$,
\begin{equation}
E_\mathrm{decay} = Q - E_\mathrm{\nu} - E_\mathrm{\gamma},
\end{equation}
since neutrinos escape the source without depositing heat, and shielding for photons is typically impractical, especially for microbatteries. Because $P_{\text{th}}$ is inversely proportional to $\thalf$, there is a tradeoff between power density and operational lifetime.

\begin{figure*}[tbph!]
\centering
\includegraphics[width=2\columnwidth]{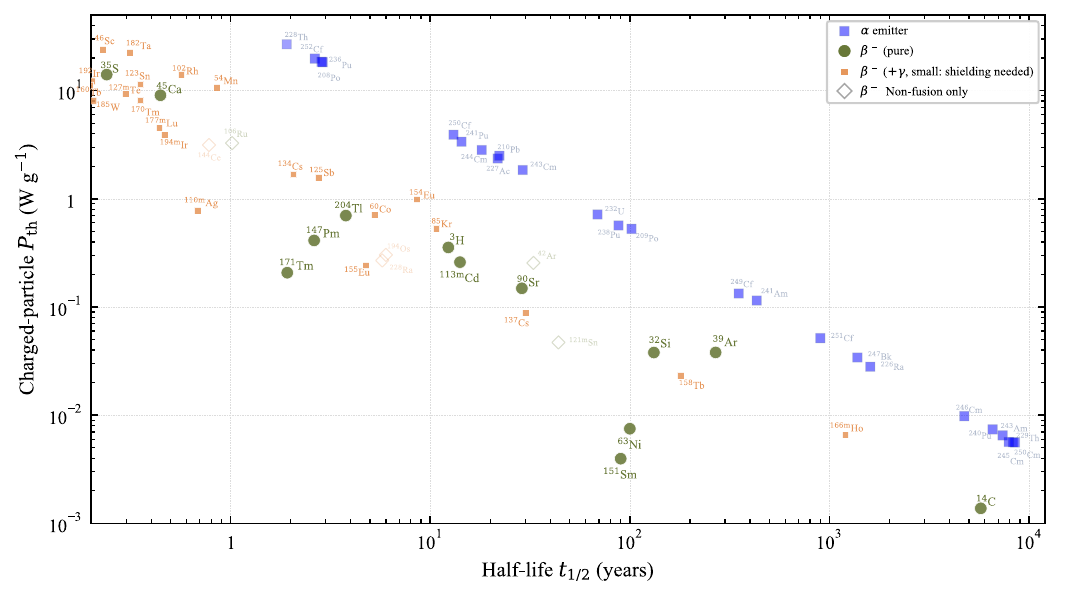}
\caption{Specific thermal power versus half-life for all nuclear battery candidate isotopes. Filled squares: fusion-producible candidates (area scales with $\freact$), coloured by decay mode: green for pure $\betam$ (no/negligible gamma), orange for $\betam$ with significant gamma. Open diamonds: non-fusion-only production pathways for $\betam$. We include $\alpha$ emitters for reference. Dashed line: ${}^{238}$Pu reference (0.57~W\,g$^{-1}$).}
\label{fig:power}
\end{figure*}

\Cref{fig:power} maps the landscape of nuclear battery radioisotopes in the specific-thermal-power/half-life plane, classified by main decay mode. Fusion-producible isotopes span the space, dominated by $\betam$ emitters from ${}^{60}$Co at 17.9~W\,g$^{-1}$ (too gamma-intense for portable use) down to ${}^{63}$Ni at 0.006~W\,g$^{-1}$ with a 100-year half-life. The pure $\betam$ emitters (green) are the most attractive for betavoltaic cells: radioisotopes such as ${}^{147}$Pm, ${}^{3}$H, ${}^{63}$Ni, ${}^{39}$Ar, and ${}^{32}$Si all have negligible gamma emission. The established actinide RTG fuels such as ${}^{238}$Pu, ${}^{241}$Am, ${}^{244}$Cm appear as blue squares - producing these alpha emitters with fusion neutrons is considered in future work. Across all nuclear battery radioisotopes, $\betam$ emitters typically have at least ten times lower specific power than alpha emitters, although their shielding requirements can be lower.

Importantly, several of these pure $\betam$ emitters either have no thermal-neutron production route at all (${}^{39}$Ar and ${}^{32}$Si) or have vanishingly low cross section until neutron energy is sufficiently high (${}^{63}$Ni). This means that fusion neutrons can uniquely produce them at high quantity at high specific activity.

\begin{figure*}[t!]
\centering
\includegraphics[width=2\columnwidth]{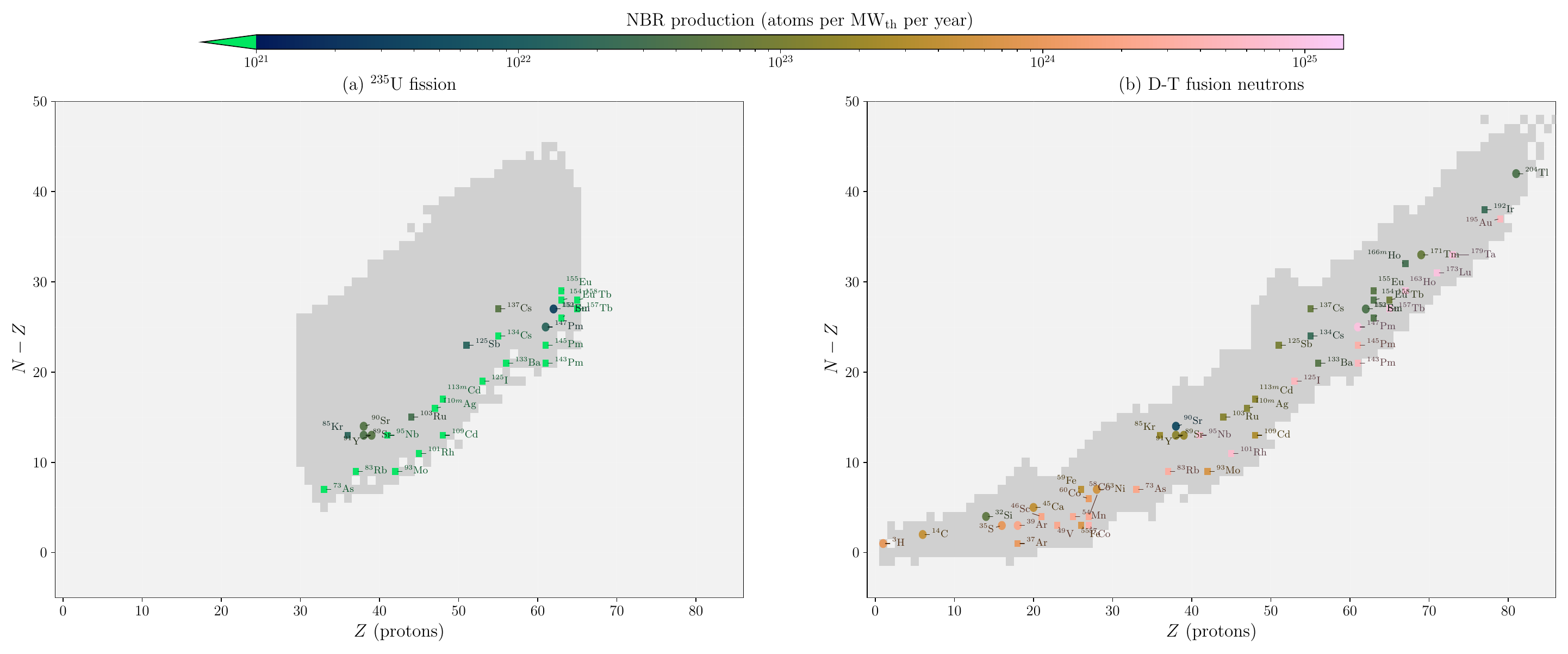}
\caption{Chart of nuclides showing production yields (atoms per MW$_{\text{th}}$ per year) for (a)~${}^{235}$U thermal fission fragments and (b)~D-T fusion neutron transmutation. Nuclear battery radioisotopes are highlighted with coloured boxes.}
\label{fig:nz}
\end{figure*}

\Cref{fig:nz} makes this point visually on the chart of nuclides. Fission products from ${}^{235}$U cluster in two mass peaks near $A \approx 95$ and $A \approx 137$, producing ${}^{90}$Sr and ${}^{137}$Cs in large quantities but missing most of the light pure $\betam$ emitters. Fusion neutrons, by contrast, can reach isotopes across the entire chart through threshold reactions on stable feedstocks.

\section{Fusion production pathways and yields} \label{sec:fusion_yields}

For each candidate isotope, we characterize the production efficiency by the \textit{neutron conversion fraction} $\freact$, the fraction of D-T source neutrons that produce the desired radioisotope. We compute $\freact$ using OpenMC Monte Carlo neutron transport simulations~\cite{openmc} with ENDF/B-VIII.0 cross-sections, modelling a 20~cm thick slab of pure feedstock irradiated by a 14.1~MeV planar source. This geometry captures the competition between the production reaction and all other neutron interactions (elastic and inelastic scattering, absorption, etc). The total annual production in atoms per year per GW of D-T power under fast-neutron irradiation is
\begin{equation}
N_{\text{product}} = \freact \dot{N}_{\text{DT}} T_\mathrm{year} ,
\end{equation}
where $\dot{N}_{\text{DT}} = 3.55 \cdot 10^{20}$~s$^{-1}$ is the number of D-T fusion reactions per second per GW of D-T thermal power, at $T_\mathrm{year} = 3.15 \cdot 10^{7}$ seconds.

We mainly consider production pathways where the product differs in atomic number from the feedstock ($\Delta Z \neq 0$). This is essential because it allows the product to be chemically separated from unconverted feedstock, yielding material with the high specific activity that nuclear batteries demand. Same-element routes ($\Delta Z = 0$) are excluded since the product and feedstock cannot be easily separated, resulting in dilute, low-activity material that is less useful for batteries - speculatively, only in the very high neutron-flux regime can a sufficient fraction of the irradiated feedstock be transmuted leading to higher specific activity - see \Cref{sec:deltaz0_n2n} for a short discussion.

\begin{figure*}[tbp!]
\centering
\includegraphics[width=2\columnwidth]{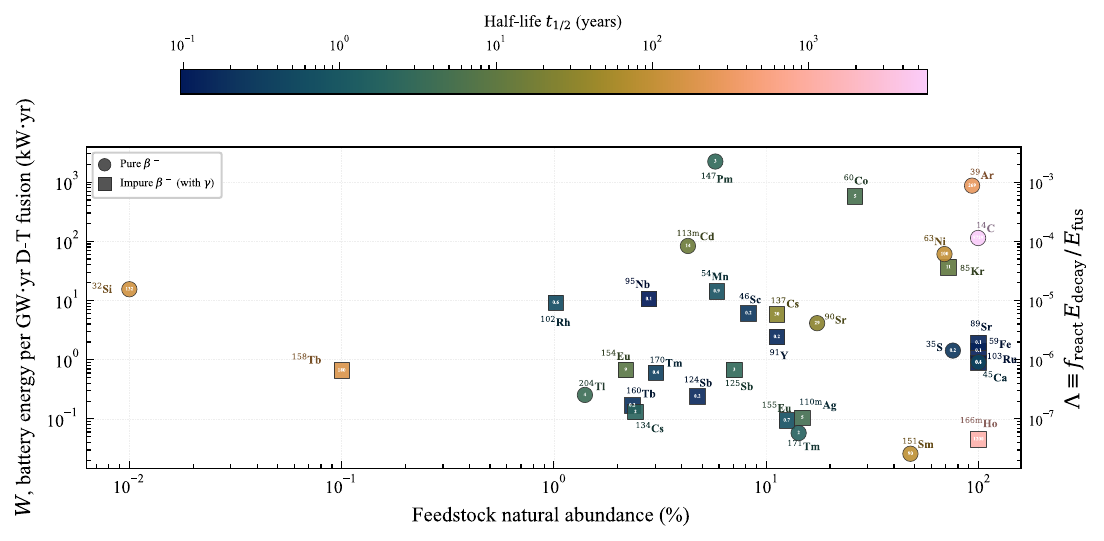}
\caption{Total battery thermal energy produced per GW-year of D-T fusion power (left axis, kW$\cdot$yr) and $\Lambda$ (right y-axis) assuming 100\% enriched feedstock, versus feedstock natural abundance. Marker shape indicates neutron economy: squares for multiplying, circles for non-multiplying reactions. Fill color maps to half-life (colorbar, top). Upper-right corner represents the most favorable candidates: high energy yield from abundant, inexpensive feedstocks. Isotopes with radioactive feedstocks are excluded.}
\label{fig:yield}
\end{figure*}

Where multiple pathways exist to the same isotope~\cite{Parisi2025}, we generally report the highest-yield route. \Cref{tab:battery} lists the key candidates; \Cref{fig:yield} ranks them by energy yield versus feedstock abundance.

\begin{table*}[htbp]
\centering
\caption{Nuclear battery isotopes producible via 14~MeV fusion neutrons, ranked by neutron conversion fraction $\freact$~n\,cm$^{-2}$\,s$^{-1}$. Production rates are for 100\% at enriched feedstock.}
\label{tab:battery}
\footnotesize
\setlength{\tabcolsep}{2pt}
\begin{tabular}{@{}lrcccccccccc@{}}
\toprule
Isotope & $\thalf$ & Decay & $\substack{E_{\betam,\max}\\\text{(keV)}}$ & $\substack{P_{\text{th}}\\\text{(W\,g$^{-1}$)}}$ & $\substack{\freact\\\text{(\%)}}$ & $\substack{\text{kg}\\\text{GW$^{-1}$yr$^{-1}$}}$ & $\substack{\text{Ci}\\\text{GW$^{-1}$yr$^{-1}$}}$ & $\Lambda$ & $\substack{\text{Feedstock}\\\text{(abund.)}}$ & Rxn & Pure $\betam$ branch  \\
\midrule
$^{147}$Pm & 2.62\,y & $\betam$ & 225 & 0.41 & 52.1 & 1{,}424 & $1.3{\cdot}10^{9}$ & $2.2{\cdot}10^{-3}$ & $^{148}$Nd (5.76\%) & \ntn & 1.00 \\[2pt]
$^{3}$H$^{\ddagger}$ & 12.3\,y & $\betam$ & 18.6 & 0.36 & ${\sim}10$ & 6 & $5.4{\cdot}10^{7}$ & $3.5{\cdot}10^{-5}$ & $^{6,7}$Li (100\%) & TBR & 1.00 \\ [2pt]
$^{60}$Co$^{\dagger}$ & 5.27\,y & $\betam$ & 318 & 0.71 & 9.5 & 106 & $1.2{\cdot}10^{8}$ & $5.7{\cdot}10^{-4}$ & $^{60}$Ni (26.2\%) & \np & 0.00 \\[2pt]
$^{39}$Ar & 269\,y & $\betam$ & 565 & 0.038 & 8.1 & 59 & $2.0{\cdot}10^{6}$ & $8.7{\cdot}10^{-4}$ & $^{39}$K (93.3\%) & \np & 1.00 \\[2pt]
$^{63}$Ni & 100\,y & $\betam$ & 67 & 0.0075 & 4.8 & 56 & $3.2{\cdot}10^{6}$ & $6.1{\cdot}10^{-5}$ & $^{63}$Cu (69.2\%) & \np & 1.00 \\[2pt]
$^{14}$C & 5{,}730\,y & $\betam$ & 156 & 0.0014 & 3.8 & 10 & $4.5{\cdot}10^{4}$ & $1.1{\cdot}10^{-4}$ & $^{14}$N (99.6\%) & \np & 1.00 \\[2pt]
$^{85}$Kr & 10.8\,y & $\betam$ & 687 & 0.53 & 0.28 & 4 & $1.7{\cdot}10^{6}$ & $3.6{\cdot}10^{-5}$ & $^{85}$Rb (72.2\%) & \np & 0.00 \\[2pt]
$^{113\text{m}}$Cd & 14.1\,y & $\betam$/IT & 586 & 0.26 & 0.75 & 16 & $3.5{\cdot}10^{6}$ & $8.3{\cdot}10^{-5}$ & $^{113}$In (4.28\%) & \np & 0.9986 \\[2pt]
$^{32}$Si & 132\,y & $\betam$ & 227 & 0.038 & 0.36 & 2 & $1.8{\cdot}10^{5}$ & $1.5{\cdot}10^{-5}$ & $^{36}$S (0.01\%) & \nna & 1.00 \\[2pt]
$^{137}$Cs$^{\S}$ & 30.1\,y & $\betam$ & 514 & 0.088 & 0.06 & 2 & $1.3{\cdot}10^{5}$ & $5.8{\cdot}10^{-6}$ & $^{137}$Ba (11.2\%) & \np & 1.00 \\[2pt]
$^{90}$Sr & 28.8\,y & $\betam$ & 546 & 0.15 & 0.04 & 1 & $9.2{\cdot}10^{4}$ & $4.1{\cdot}10^{-6}$ & $^{94}$Zr (17.4\%) & \nna & 1.00 \\
\bottomrule
\end{tabular}
\begin{flushleft}
\scriptsize $P_{\text{th}}$ values are charged-particle power only ($\langle E_\beta \rangle \approx E_{\betam,\max}/3$ for $\betam$ emitters), excluding gamma and neutrino energy.
$^{\dagger}$The listed $P_{\text{th}} = 0.71$~W\,g$^{-1}$ is beta-only; total including gammas is 17.9~W\,g$^{-1}$, but the gammas (1.17 and 1.33~MeV, combined intensity ${\approx}\,200\%$) require massive shielding.
$^{\ddagger}$Tritium is produced via excess tritium breeding from a ${}^{6}$Li/${}^{7}$Li blanket (TBR excess ${\approx}\,0.1$), not by transmutation of ${}^{3}$He.
\end{flushleft}
\end{table*}

${}^{147}$Pm has the highest yield ($\freact = 52.1\%$) under fast neutrons of any $\beta^{-}$ battery candidate considered here, produced through the neutron-multiplying chain 
\begin{equation}
{}^{148}\mathrm{Nd}\xrightarrow{\ntn}{}^{147}\mathrm{Nd}\xrightarrow{\betam}{}^{147}\mathrm{Pm}.
\end{equation}
With a beta energy of 225~keV and negligible gamma emission, it is well matched to semiconductor betavoltaic junctions \cite{flicker1964construction} and was historically used in cardiac pacemaker batteries~\cite{Prelas2016}. Various ${}^{147}$Pm-powered nuclear batteries have been built~\cite{flicker1964construction,yakubova2010nuclear,kavetskiy2011efficiency,waris2016preliminary}. The feedstock ${}^{148}$Nd (5.76\% of natural neodymium) still produces significant quantities of ${}^{147}$Pm without enrichment, and the $\ntn$ reaction returns two neutrons to the blanket, imposing negligible, if any, penalty on the tritium breeding ratio (TBR) compared with other neutron multipliers. The D-T neutron (n,2n) approach is less challenging than the fission reactor production pathway ~\cite{broderick2019reactor}, which struggles to obtain high ${}^{147}$Pm yield because of the large ${}^{147}$Pm(n,$\gamma$) cross section at lower neutron energy - a flowing fusion blanket with a ${}^{148}$Nd alloy and regular ${}^{147}$Pm extraction can maintain high yields by avoiding ${}^{147}$Nd(n,$\gamma$) and ${}^{147}$Pm(n,$\gamma$) burnup.

${}^{39}$Ar has a high yield on abundant feedstock ${}^{39}$K and a high cross section of several hundred millibarns. Its use has been suggested \cite{Jurewitsch1998radionuclide} for radioluminescent nuclear batteries~\cite{xu2020radioluminescent}, with impressive specific power given its 268 year half life. Production of ${}^{85}$Kr, another noble gas, is also relatively high on the feedstock ${}^{85}$Rb with natural abundance 72.2\%.

${}^{63}$Ni ($\thalf = 100$\,y, $E_{\betam,\max} = 67$~keV, $\freact = 4.8\%$) is a good candidate for long-lived betavoltaic cells \cite{tang2012optimization,munson2015model}. Its very low beta energy gives excellent semiconductor conversion efficiency, and the absence of gamma emission means no lattice damage accumulates over decades of operation, nor is cumbersone shielding required. It is produced from natural copper (69.2\% ${}^{63}$Cu) via ${}^{63}$Cu$\np$, a reaction whose cross section only exceeds 1mb at $E_n \simeq 1$ MeV and 10mb at $E_n \simeq 2$ MeV.

${}^{32}$Si ($\thalf = 132$\,y, $E_{\betam,\max} = 227$~keV, $\freact = 0.36\%$) is another pure $\betam$ emitter, produced via ${}^{36}$S$\nna$. Despite the appealing properties of ${}^{32}$Si, the natural abundance of ${}^{36}$S is prohibitively low (0.01\%).

${}^{60}$Co has a high yield on abundant feedstock ${}^{60}$Ni. However, due to ${}^{60}$Co emitting very strong gamma radiation, we generally rule it out in this analysis as a nuclear battery material.

\section{Production scale estimates}
\label{sec:scale}

For (n,2n)-produced ${}^{147}$Pm, the availability of ${}^{147}$Pm can scale with the size of the amount of deployed fusion power. Assuming a fraction $f_{{}^{148}\mathrm{Nd}}$ of fusion plants use ${}^{148}$Nd as a multiplier material, with ${}^{148}$Nd being enriched in Nd to fraction $\epsilon_{{}^{148}\mathrm{Nd}}$, the annual production of ${}^{147}$Pm in tons is
\begin{equation}
\begin{aligned}
& \dot{M}_{{}^{147}\mathrm{Pm}} = f_\mathrm{react} f_{{}^{148}\mathrm{Nd}} \epsilon_{{}^{148}\mathrm{Nd}} \frac{P_\mathrm{fus}}{E_\mathrm{fus}} M_{{}^{147}\mathrm{Pm}} T_\mathrm{year} \\
& \approx 2.7 f_\mathrm{react} f_{{}^{148}\mathrm{Nd}} \epsilon_{{}^{148}\mathrm{Nd}} P_\mathrm{fus}^\mathrm{GW} \;\; [\mathrm{t / yr}].
\end{aligned}
\end{equation}
Here, $P_\mathrm{fus}$ and $P_\mathrm{fus}^\mathrm{GW}$ are the total installed fusion capacity in watts and gigawatts, $E_\mathrm{fus} = 2.8 \cdot 10^{-12}$J is the energy released per fusion reaction, and $M_{{}^{147}\mathrm{Pm}}$ is the mass of a ${}^{147}$Pm atom in tons.

The total useful energy emitted over the NBR lifetime per gigawatt year of fusion power is
\begin{equation}
W \equiv \dot{M} E_\mathrm{decay}.
\end{equation}
This can be written as 
\begin{equation}
W = \Lambda f_\mathrm{target} \epsilon_\mathrm{target} P_\mathrm{fus} T_\mathrm{year} M_\mathrm{target},
\end{equation}
where $f_\mathrm{target}$, $\epsilon_\mathrm{target}$, and $M_\mathrm{target}$ are general forms of $f_{{}^{148}\mathrm{Nd}}$, $\epsilon_{{}^{148}\mathrm{Nd}}$, and $M_{{}^{147}\mathrm{Pm}}$, and $\Lambda$ is the fraction of $E_\mathrm{fus} $ that is released by the radioisotope,
\begin{equation}
\Lambda \equiv \frac{f_\mathrm{react} E_\mathrm{decay} }{E_\mathrm{fus} } .
\end{equation}
$\Lambda$ can be thought of as an `efficiency factor' - the fraction of fusion deployment energy that is deployed in nuclear batteries. \Cref{fig:yield} shows $\Lambda$ (right y-axis) for a range of fuels - $\Lambda$ represents the maximum theoretical available fraction of the fusion energy capacity that can be deployed in nuclear batteries, and is highest at $\Lambda \simeq 2 \cdot 10^{-3}$ for ${}^{147}$Pm.

\Cref{fig:yield} shows the total battery thermal energy produced per GW-year of D-T fusion power versus feedstock natural abundance. The upper-right corner represents the most favourable candidates: high energy yield from abundant, inexpensive feedstocks requiring no isotopic enrichment. ${}^{60}$Co from natural Ni (26\%) and ${}^{14}$C from natural N (99.6\%at ${}^{14}$N) lead in raw energy output but are limited by intense gammas and very low specific power, respectively. Among practical pure-$\betam$ candidates, ${}^{147}$Pm (from 5.76\% ${}^{148}$Nd) produces ${\sim}\,2.2$~MW$\cdot$yr of battery energy per GW$\cdot$yr of fusion, ${}^{63}$Ni ${\sim}\,0.06$~MW$\cdot$yr, and ${}^{39}$Ar ${\sim}\,0.87$~MW$\cdot$yr.

\begin{figure}[bp!]
\centering
\includegraphics[width=\columnwidth]{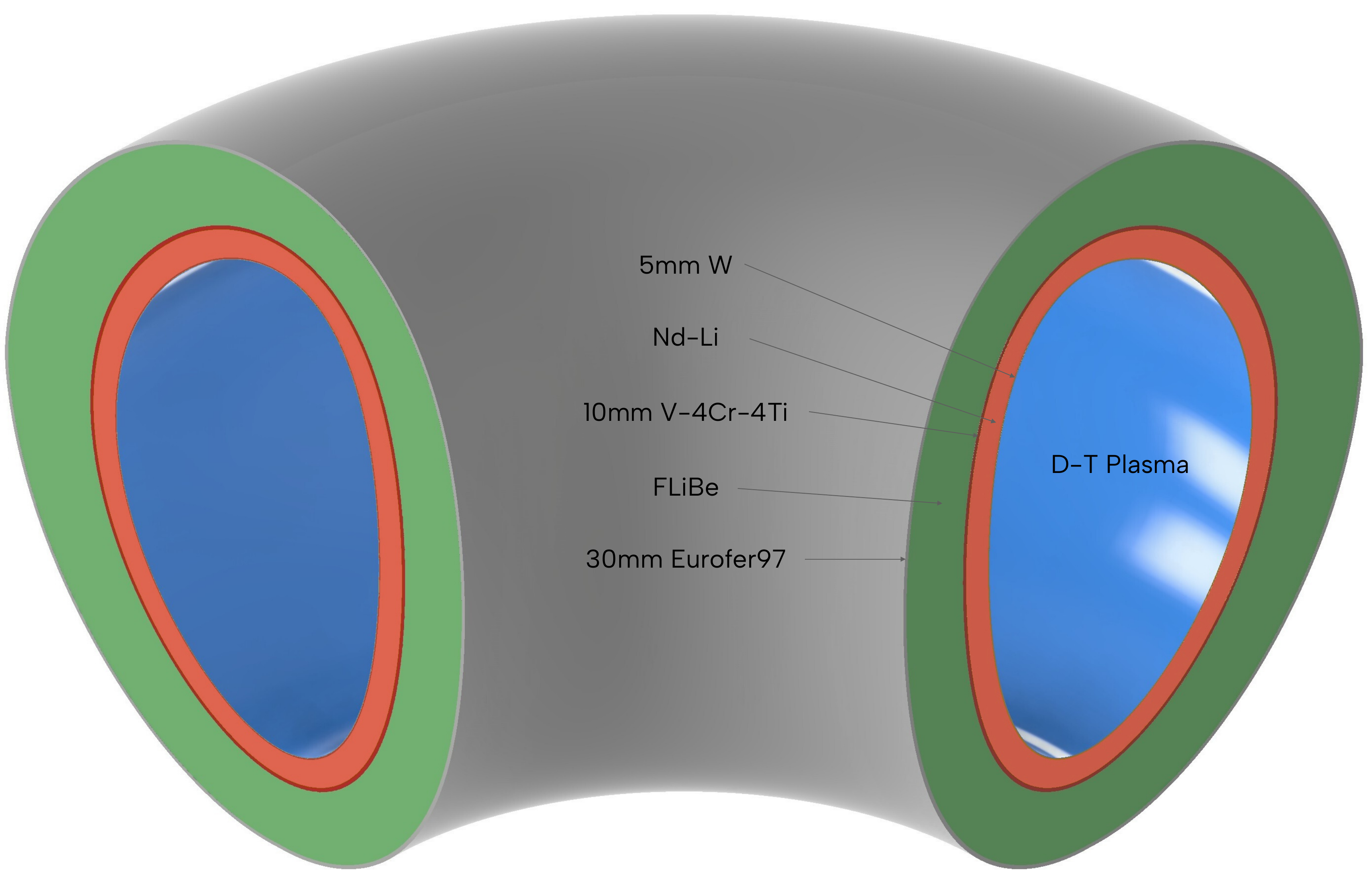}
\caption{Tokamak blanket geometry used in the neutronics simulations. From plasma outward: tungsten first wall (5~mm), V-4Cr-4Ti structural layer (10~mm), enriched ${}^{148}$Nd\,+\,Li channel (variable thickness, coral), V-4Cr-4Ti structural layer (30~mm), FLiBe (Li$_2$BeF$_4$) tritium breeding blanket (green), and EUROFER97 outer wall (30~mm, grey). The channel thickness and Nd/Li ratio are varied in the scan of \Cref{fig:blanket_scan}.}
\label{fig:simulation_setup}
\end{figure}

\begin{figure*}[tbp!]
\centering
\includegraphics[width=1.8\columnwidth]{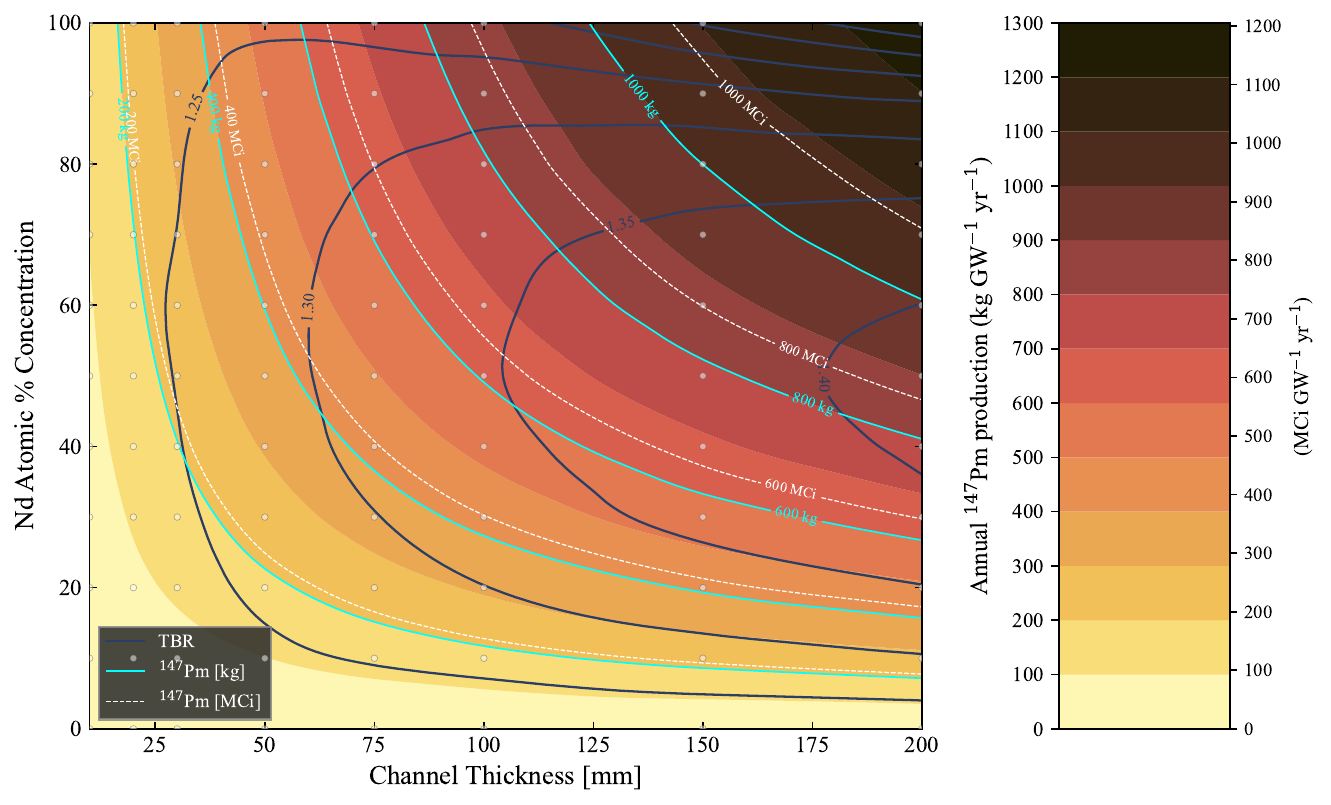}
\caption{Annual ${}^{147}$Pm production (kg\,GW$^{-1}$\,yr$^{-1}$ and MCi\,GW$^{-1}$\,yr$^{-1}$) and tritium breeding ratio (TBR, black contour lines) as a function of channel thickness and Nd atomic concentration in the channel layer. The channel contains 90\%\,at enriched ${}^{148}$Nd mixed with 90\%\,at enriched ${}^{6}$Li, embedded in the tokamak blanket geometry of \cref{fig:simulation_setup}. The FLiBe breeder behind the channel maintains TBR~$> 1$ across the entire parameter space, with TBR decreasing as the channel becomes thicker and more Nd-rich. Cyan contour lines indicate ${}^{147}$Pm production in kg\,GW$^{-1}$\,yr$^{-1}$. Computed with OpenMC~\cite{openmc} using ENDF/B-VIII.0 cross-sections.}
\label{fig:blanket_scan}
\end{figure*}

The production columns in \Cref{tab:battery} give the annual output from a single GW-class D-T fusion plant, computed from OpenMC transport simulations of a 20~cm thick feedstock slab irradiated by 14.1~MeV neutrons. A single GW${}_\mathrm{th}$ plant could produce ${\sim}\,1{,}400$~kg\,yr$^{-1}$ (${\sim}\,1.3 \cdot 10^{9}$~Ci) of ${}^{147}$Pm, an isotope for which no large-scale supply currently exists, along with ${\sim}\,56$~kg\,yr$^{-1}$ of ${}^{63}$Ni and ${\sim}\,59$~kg\,yr$^{-1}$ of ${}^{39}$Ar.

\section{Neutron economy and tritium breeding compatibility}
\label{sec:neutron}

A critical question for any co-production scheme in a D-T fusion blanket is whether feedstock transmutation competes with tritium breeding~\cite{Abdou2021}. For a material to be a useful blanket constituent, it must function as a neutron multiplier, a moderator, or a tritium breeder. None of the feedstocks studied here except ${}^{6,7}$Li have tritium breeding properties, so blanket compatibility depends on whether the feedstock is a good multiplier or moderator.

Several of the higher-$Z$ feedstocks are acceptable neutron multipliers, provided their $\ntn$ cross section is sufficiently high ($\gtrsim 1.5$~b at 14~MeV) and their (n,$\gamma)$ cross section is low. Isotopes such as ${}^{148}$Nd, ${}^{151}$Eu, and ${}^{94}$Zr satisfy these criteria and might replace conventional multiplier materials (Be, Pb) in part of the blanket while simultaneously producing NBRs. ${}^{148}$Nd is particularly attractive: its $\ntn$ reaction both multiplies neutrons and produces the highest-yield battery candidate, ${}^{147}$Pm.

Similar to the setup in~\cite{Rutkowski2025}, We perform OpenMC simulations of a two-layer blanket system (see \Cref{fig:simulation_setup}), with 90\%at enriched ${}^{148}$Nd, 90\%at enriched ${}^{6}$Li, and varying Nd fraction in the first layer. The ${}^{148}$Nd layer is varied in thickness, with the overall radial build fixed width. In \Cref{fig:blanket_scan}, we plot ${}^{147}$Pm annual production and TBR. There exist many design points where TBR$\gtrsim 1.1$ --- an approximate required value for tritium self-sufficiency --- and ${}^{147}$Pm production exceeds 1000 kg per year.

Other feedstocks are neither good multipliers nor effective moderators. Low-$Z$ targets such as ${}^{14}$N and ${}^{36}$S have small $\ntn$ cross sections and would reduce the TBR if used in large volume fractions. For these materials, only a fraction of the blanket volume can be allocated to feedstock, with the remainder compensating via dedicated multiplier and breeder zones. Integrated blanket neutronic simulations would be needed to quantify the achievable TBR for each feedstock configuration.

We perform OpenMC depletion simulations to determine the presence of long-lived radioisotopes after 30 years of operating with a 100mm enriched ${}^{148}$Nd blanket enclosed by a FLiBe breeding layer. The two longest-lived radioisotopes are ${}^{147}$Pm, and ${}^{151}$Sm, the latter has multiple production pathways. ${}^{151}$Sm is also a betavoltaic fuel, and can be collected separately from ${}^{147}$Pm. Separate harvesting of ${}^{147}$Pm and ${}^{151}$Sm allows the remaining blanket material to be classified as Class A low-level waste about six years after irradiation has ended. More details are provided in Appendix \ref{app:tokamak_PmSm}. 

\section{$\Delta Z = 0$ pathways}
\label{sec:deltaz0_n2n}

We briefly discuss producing material bearing NBRs produced by (n,2n) reactions that do not change the proton number $Z$ from the irradiated feedstock. Because chemical methods cannot be used to extract NBR from the feedstock, and because isotopic separation for large quantities of radioactive material is challenging, the full irradiated feedstock will be considered as the nuclear battery material. These routes are typically attractive only at fluxes where a substantial fraction of the feedstock is converted so that the specific power is sufficiently high. The advantage is that because (n,2n) cross sections are typically at least hundreds of millibarns, the production rate is high.

\begin{figure}[b!]
\centering
\includegraphics[width=\columnwidth]{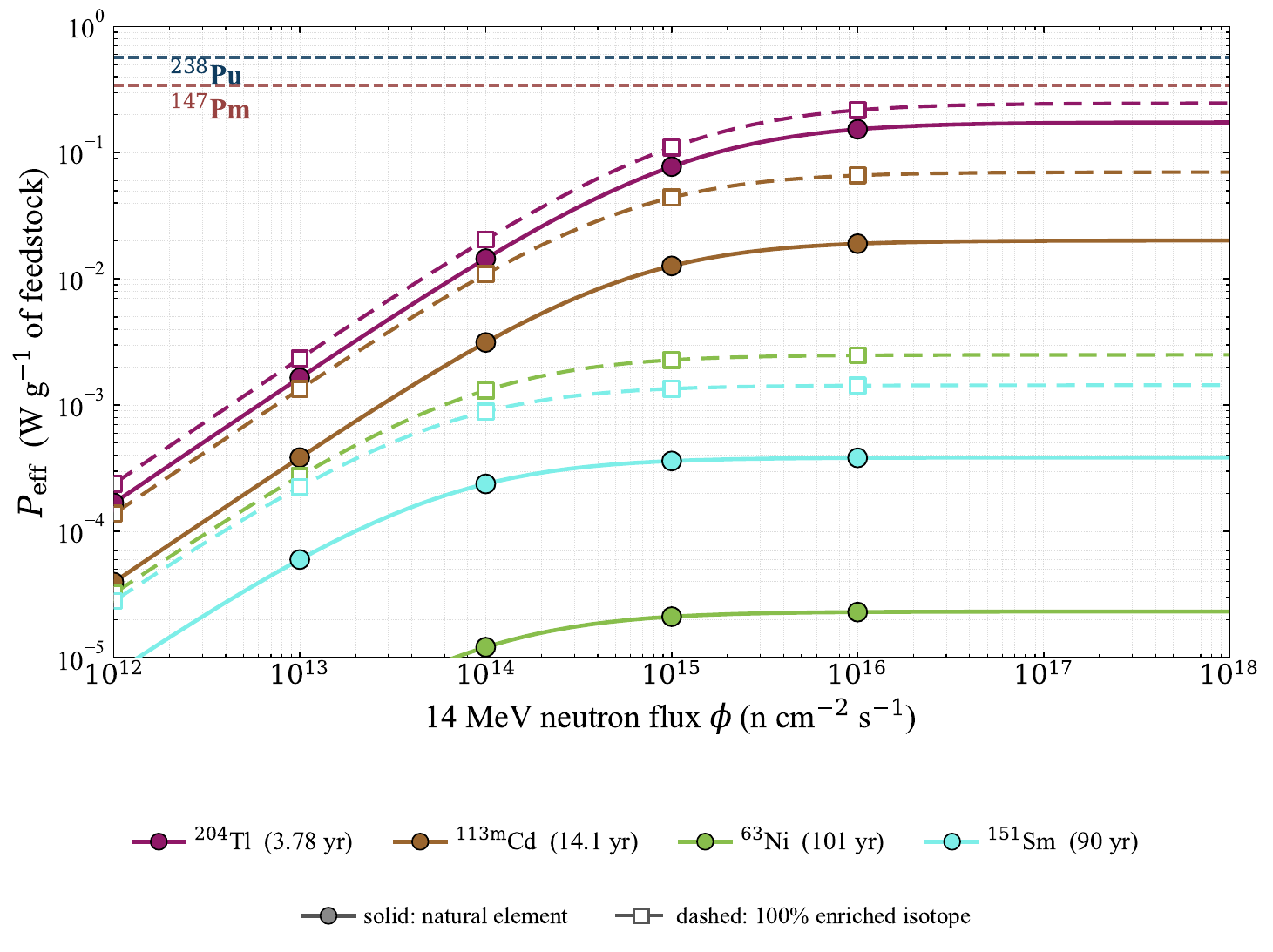}
\caption{Charged-particle specific power versus 14\,MeV neutron flux for the pure-$\betam$ $\Delta Z = 0$ (n,2n) candidates. Solid lines and filled circles: natural feedstock element. Dashed lines and open squares: 100\% enriched feedstock isotope. Dashed horizontal references mark the pure-product ${}^{238}$Pu and ${}^{147}$Pm specific power.}
\label{fig:deltaz0_pvf}
\end{figure}

Restricting to pure $\betam$ emitters with useful battery half-lives leaves four single-step candidates: ${}^{205}$Tl${\ntn}{}^{204}$Tl (3.8\,yr, $E_\beta^{\max}=764$\,keV), ${}^{114}$Cd${\ntn}{}^{113\mathrm{m}}$Cd (14.1\,yr, $585$\,keV; ${\sim}45\%$ isomer branch), ${}^{152}$Sm${\ntn}{}^{151}$Sm (90\,yr, $76$\,keV), and ${}^{64}$Ni${\ntn}{}^{63}$Ni (101\,yr, $67$\,keV). \Cref{fig:deltaz0_pvf} plots the charged-particle specific power versus 14 MeV neutron flux for each candidate at natural feedstock abundance and at 100\% enrichment. 

Specific power curves saturate when the production rate $\sigma \phi$ balances the decay rate $\lambda$, giving the saturation neutron flux $\phi \sim \lambda / \sigma$. At natural feedstock abundance only ${}^{204}$Tl approaches ${}^{147}$Pm in specific power; enrichment is essential for ${}^{63}$Ni (only $0.93\%$ of natural Ni is ${}^{64}$Ni). Long-lived nuclear isomers such as ${}^{180 \mathrm{m} }\mathrm{Ta}$ might also be produced at scale via (n,2n) reactions, although there exists no known way to reliably trigger \cite{palffy2007isomer,arnquist2023constraints,ding2026isomer} ${}^{180 \mathrm{m} }\mathrm{Ta}$ and release the energy via electrons.

\section{Value per Neutron} \label{sec:vpn}

An essential part of determining whether NBRs are worth producing is whether their production increases the value per fusion neutron \cite{parisi2026neutronvalue}. The value per neutron of ${}^{147}$Pm produced in fusion blankets can be estimated from the prices of existing nuclear batteries. Here, we use the price of a ${}^{238}$Pu RTG generator as an example - assuming an electrical conversion efficiency of $\eta_{{}^{238}\mathrm{Pu}}=$7\%, $E_\mathrm{decay} = 8.95 \cdot 10^{-13}$ J, and a molar mass of 238g. This gives an electric stored energy per gram of fuel of
\begin{equation}
    \epsilon_\mathrm{kJ}^\mathrm{fuel} = \eta_{{}^{238}\mathrm{Pu}} \frac{E_\mathrm{decay} N_\mathrm{A}}{m_{{}^{238}\mathrm{Pu}}} = 1.58 \cdot 10^{5} \; \mathrm{kJ} / \mathrm{g}.
\end{equation}
Assuming that the structure mass is 1000 times heavier than the fuel itself, this gives the total battery electric stored energy per gram
\begin{equation}
    \epsilon_\mathrm{kJ}^\mathrm{battery} = 1.58 \cdot 10^{2} \; \mathrm{kJ} / \mathrm{g}.
\end{equation}
Assuming a fuel price of $C_{{}^{238}\mathrm{Pu}} =$\$10,000/g, the cost per kJ is
\begin{equation}
    c_{{}^{238}\mathrm{Pu}} = \$ 6.3 / \mathrm{kJ}.
\end{equation}
Assuming that the cost of electricity from a baseload power source is \$60/$\mathrm{MWh}_\mathrm{e}$, the grid cost per kJ is
\begin{equation}
    c_\mathrm{grid} = \$ 1.7 \cdot 10^{-5} / \mathrm{kJ}.
\end{equation}
Therefore, the price premium per unit energy of a ${}^{238}$Pu battery over grid energy is
\begin{equation}
    \mathcal{P} = \frac{c_{{}^{238}\mathrm{Pu}}}{c_\mathrm{grid}} = 3.8 \cdot 10^{5}.
\end{equation}
We can now calculate the value per neutron of ${}^{147}$Pm in a nuclear battery by using $\mathcal{P}$. The amount of usable electric energy in a nuclear battery generated by decay of a single ${}^{147}$Pm atom is
\begin{equation}
    E_{{}^{147}\mathrm{Pm}} = E_\mathrm{decay} \eta_{{}^{147}\mathrm{Pm}} = 6.0 \cdot 10^{-19} \; \mathrm{kJ},
\end{equation}
where we used $E_\mathrm{decay} = 9.9 \cdot 10^{-18}$ kJ and $\eta_{{}^{147}\mathrm{Pm}} = 0.06$. If the revenue generated per neutron of electricity is $v_\mathrm{n}^\mathrm{elec} = \$10^{-20}$. Therefore, the revenue per neutron generated by ${}^{147}$Pm production is 
\begin{equation}
    v_{{}^{147}\mathrm{Pm}} = v_\mathrm{n}^\mathrm{elec} f_\mathrm{react} \frac{E_{{}^{147}\mathrm{Pm}}}{\eta_\mathrm{th} E_\mathrm{n}} \frac{ \mathcal{P}}{\mathcal{M}},
    \label{eq:v_per_neutron}
\end{equation}
where $\eta_\mathrm{th} = 0.3$ is the thermal-to-electric conversion efficiency in a fusion power plant, $E_\mathrm{n} = 14.5$ MeV is the D-T neutron energy, and $\mathcal{M}$ is the ratio of the battery structure cost to the fuel. Here, we assume $\mathcal{M} = 100$.

\begin{figure}[t!]
\centering
\includegraphics[width=\columnwidth]{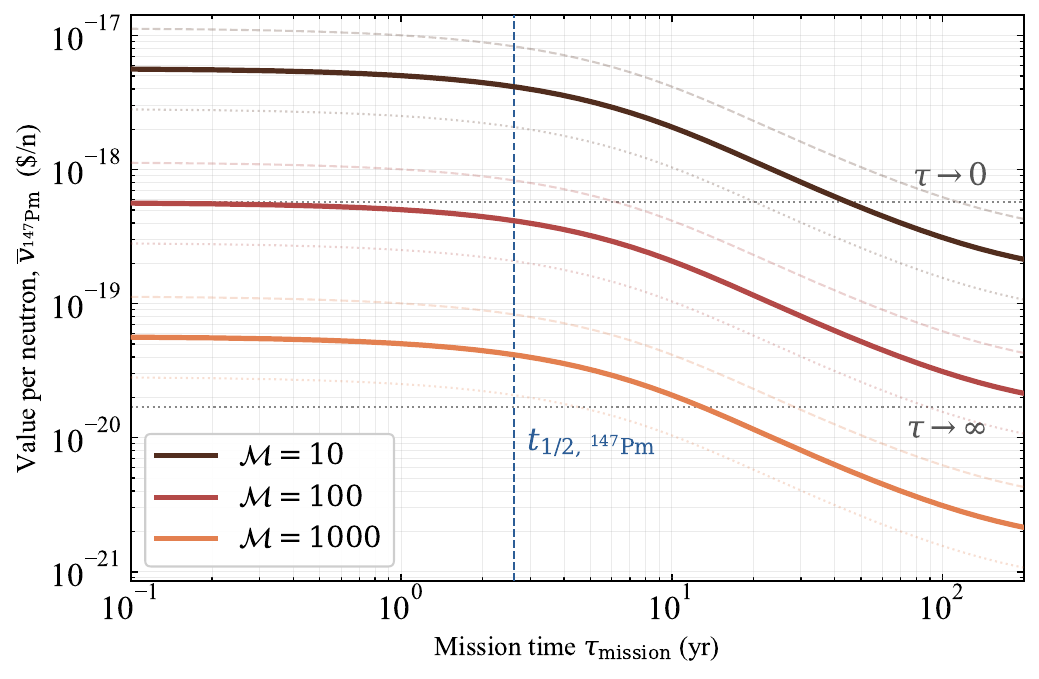}
\caption{Value per neutron (see \Cref{eq:v_neutron_final}) for ${}^{147}$Pm produced in a fusion blanket assuming $f_\mathrm{react} = 0.52$.}
\label{fig:value_per_neutron}
\end{figure}

There is a further enhancement to the value per neutron in \Cref{eq:v_per_neutron} for shorter battery missions of time $\tau_\mathrm{mission}$, which is due to the much shorter half-life of ${}^{147}$Pm compared with ${}^{238}$Pu. The enhancement factor is
\begin{equation}
    \mathcal{E}(\tau_\mathrm{mission}) = \frac{1-\exp\left(-\ln 2 \tau_\mathrm{mission}/t_{1/2,{}^{147}\mathrm{Pm}} \right)}{1-\exp\left(-\ln 2 \tau_\mathrm{mission}/t_{1/2,{}^{238}\mathrm{Pu}} \right)}.
\end{equation}
Therefore, in a simple model the value per neutron also depends on the relative mission time of ${}^{147}$Pm compared with ${}^{238}$Pu, giving the value per neutron
\begin{equation}
    \overline{v}_{{}^{147}\mathrm{Pm}} = \mathcal{E} v_{{}^{147}\mathrm{Pm}}.
    \label{eq:v_neutron_final}
\end{equation}
The value per fusion neutron of producing ${}^{147}$Pm is shown in \Cref{fig:value_per_neutron} for different mission times. There are significant sources of uncertainty in \Cref{eq:v_neutron_final}. Most crucially, we used the price premium $\mathcal{P}$ that we calculated from an alpha emitter and applied it to a beta emitter. Given that alpha and beta emitters have very different use cases, this assumption may be unfounded - if beta emitters have a higher $\mathcal{P}$ value than ${}^{238}$Pu for example, or if $\mathcal{M}$ is lower for beta emitters, the value per neutron could be higher. Lower $\mathcal{M}$ values could be achieved if for example, much reduced shielding is required compared with alpha emitters.

For reference, the value per neutron of electricity is roughly \$$10^{-20}$/neutron~\cite{parisi2026neutronvalue}, which is tens to hundreds of times lower than the typical value per neutron of ${}^{147}$Pm in \Cref{fig:value_per_neutron}. Since electricity can be co-produced with ${}^{147}$Pm, an electricity-generating fusion power plant could significantly increase its economic value by co-producing ${}^{147}$Pm, provided there is a sufficiently large market for ${}^{147}$Pm.

\section{Discussion} \label{sec:discussion}

14 MeV D-T fusion neutrons can produce a broad range of nuclear battery radioisotopes (NBRs) spanning the specific-power/half-life space. The most promising NBR with the highest-yield pathway is production of the pure $\betam$ emitter ${}^{147}$Pm, with a production rate of ${\sim}\,1$~tonnes\, GWyr$^{-1}$. Other promising NBRs include ${\sim}\,56$~kg\,GWyr$^{-1}$ of ${}^{63}$Ni, and ${\sim}\,59$~kg\,GWyr$^{-1}$ of ${}^{39}$Ar. Tritium from excess tritium breeding in ${}^{6,7}$Li blankets provides an additional betavoltaic fuel. Many of these isotopes, including ${}^{147}$Pm, ${}^{63}$Ni, ${}^{39}$Ar, ${}^{32}$Si, and ${}^{14}$C, have thermal-neutron production routes orders of magnitude less efficient than with fusion neutrons. By expanding the availability of pure $\betam$ radioisotopes, the fusion industry could open a path to broad commercialization of nuclear batteries using $\betam$ emitters. Production of $\alpha$ emitters with fusion neutrons will be discussed in a future work.

All of these production figures are calculated under idealized scenarios. While we have identified nuclear battery radioisotopes that can be produced at scale, for any given radioisotope, extensive design work is required to optimize production and minimize losses within the constraints of fusion blanket systems.

\section{Acknowledgements}

We are grateful for conversations with A. Rutkowski, S. Chen, and A. Vernon.

\section*{Data availability}

Data used in this study will be made publicly available upon publication.

\bibliographystyle{unsrtnat}
\bibliography{references}

\begin{thebibliography}{48}
\providecommand{\natexlab}[1]{#1}
\providecommand{\url}[1]{\texttt{#1}}
\expandafter\ifx\csname urlstyle\endcsname\relax
  \providecommand{\doi}[1]{doi: #1}\else
  \providecommand{\doi}{doi: \begingroup \urlstyle{rm}\Url}\fi

\bibitem[Prelas et~al.(2014)Prelas, Weaver, Watermann, Lukosi, Schott, and
  Wisber]{Prelas2016}
M.~A. Prelas, C.~L. Weaver, M.~L. Watermann, E.~D. Lukosi, R.~J. Schott, and
  D.~A. Wisber.
\newblock A review of nuclear batteries.
\newblock \emph{Prog. Nucl. Energy}, 75:\penalty0 117--148, 2014.
\newblock URL \url{https://doi.org/10.1016/j.pnucene.2014.04.007}.

\bibitem[Olsen et~al.(2012)Olsen, Cabauy, and Elkind]{Olsen2012}
Larry~C Olsen, Peter Cabauy, and Bret~J Elkind.
\newblock Betavoltaic power sources.
\newblock \emph{Physics today}, 65\penalty0 (12):\penalty0 35--38, 2012.

\bibitem[O'brien et~al.(2008)O'brien, Ambrosi, Bannister, Howe, and
  Atkinson]{OBrien2008}
RC~O'brien, RM~Ambrosi, NP~Bannister, SD~Howe, and Helen~V Atkinson.
\newblock Safe radioisotope thermoelectric generators and heat sources for
  space applications.
\newblock \emph{Journal of Nuclear Materials}, 377\penalty0 (3):\penalty0
  506--521, 2008.

\bibitem[Dustin and Borrelli(2021)]{Dustin2021}
J~Seth Dustin and Robert~A Borrelli.
\newblock Assessment of alternative radionuclides for use in a radioisotope
  thermoelectric generator.
\newblock \emph{Nuclear Engineering and Design}, 385:\penalty0 111475, 2021.

\bibitem[Krasnov and Legotin(2020)]{Krasnov2021}
AA~Krasnov and SA~Legotin.
\newblock Advances in the development of betavoltaic power sources (a review).
\newblock \emph{Instruments and Experimental Techniques}, 63\penalty0
  (4):\penalty0 437--452, 2020.

\bibitem[LaPotin et~al.(2022)LaPotin, Schulte, Steiner, Buznitsky, Kelsall,
  Friedman, Tervo, France, Young, Rohskopf, et~al.]{Lapotin2022}
Alina LaPotin, Kevin~L Schulte, Myles~A Steiner, Kyle Buznitsky, Colin~C
  Kelsall, Daniel~J Friedman, Eric~J Tervo, Ryan~M France, Michelle~R Young,
  Andrew Rohskopf, et~al.
\newblock Thermophotovoltaic efficiency of 40\%.
\newblock \emph{Nature}, 604\penalty0 (7905):\penalty0 287--291, 2022.

\bibitem[Wang et~al.(2019)Wang, Han, Zhang, Li, Li, Zhao, and Wu]{Wang2019}
Xiaoyu Wang, Yuncheng Han, Jiachen Zhang, Ziwei Li, Taosheng Li, Xueyan Zhao,
  and Yican Wu.
\newblock The design of a direct charge nuclear battery with high energy
  conversion efficiency.
\newblock \emph{Applied Radiation and Isotopes}, 148:\penalty0 147--151, 2019.

\bibitem[Rinehart(2001)]{Rinehart2001}
Gary~H Rinehart.
\newblock Design characteristics and fabrication of radioisotope heat sources
  for space missions.
\newblock \emph{Progress in Nuclear Energy}, 39\penalty0 (3-4):\penalty0
  305--319, 2001.

\bibitem[Cataldo and Paffett(2020)]{Cataldo2020}
E.~Cataldo and G.~Paffett.
\newblock Radioisotope power systems for deep space exploration.
\newblock In \emph{Nuclear and Emerging Technologies for Space}. American
  Nuclear Society, 2020.

\bibitem[Foust(2023)]{SpaceNewsPu238}
J.~Foust.
\newblock Full-scale production of plutonium-238 still years away.
\newblock \emph{SpaceNews}, 2023.
\newblock URL
  \url{https://spacenews.com/full-scale-production-of-plutonium-238-still-years-away/}.

\bibitem[{Oak Ridge National Laboratory}(2023)]{ORNLPu238}
{Oak Ridge National Laboratory}.
\newblock {Pu-238} shipment quantity ``opens the tap'' for space missions,
  2023.
\newblock URL
  \url{https://www.ornl.gov/news/pu-238-shipment-quantity-opens-tap-space-missions}.
\newblock ORNL News.

\bibitem[{Idaho National Laboratory}(2023)]{INLPu238}
{Idaho National Laboratory}.
\newblock National labs resume plutonium production for space exploration,
  2023.
\newblock URL
  \url{https://inl.gov/feature-story/national-labs-resume-plutonium-production-for-space-exploration/}.
\newblock INL Feature Story.

\bibitem[Ambrosi et~al.(2019)]{Ambrosi2019}
R.~M. Ambrosi et~al.
\newblock European radioisotope thermoelectric generators ({RTGs}) and
  radioisotope heater units ({RHUs}) for space science and exploration.
\newblock \emph{Space Sci. Rev.}, 215:\penalty0 55, 2019.
\newblock URL \url{https://doi.org/10.1007/s11214-019-0623-9}.

\bibitem[Sheridan et~al.(2018)]{AMPPEX}
K.~Sheridan et~al.
\newblock Americium and plutonium purification by extraction (the {AMPPEX}
  process): development of a new method to separate ${}^{241}${Am} from aged
  plutonium dioxide for use in space power systems.
\newblock \emph{J. Inorg. Nucl. Chem.}, 2018.

\bibitem[Pavlov et~al.(2024)]{Ni63Production}
S.~V. Pavlov et~al.
\newblock An integrated closed loop flowsheet for production of highly enriched
  ${}^{63}${Ni} and deposition of ${}^{63}${Ni} coatings.
\newblock \emph{Radiochemistry}, 66:\penalty0 1023, 2024.
\newblock URL \url{https://doi.org/10.1134/S1066362224060018}.

\bibitem[Blanchard(2025)]{IEEESpectrum2025}
J.~Blanchard.
\newblock Nuclear batteries: energy storage for decades.
\newblock \emph{IEEE Spectrum}, August 2025.
\newblock URL \url{https://spectrum.ieee.org/nuclear-battery-revival}.

\bibitem[{University of Bristol and Arkenlight}(2024)]{Arkenlight2024}
{University of Bristol and Arkenlight}.
\newblock World's first carbon-14 diamond battery prototype demonstrated,
  December 2024.
\newblock URL
  \url{https://www.bristol.ac.uk/news/2024/december/diamond-battery-media-release.html}.
\newblock Press Release.

\bibitem[{City Labs Inc.}(2024)]{CityLabs}
{City Labs Inc.}
\newblock {NanoTritium} betavoltaic battery technology, 2024.
\newblock URL \url{https://citylabs.net/products/}.

\bibitem[{U.S. Department of Energy}(2024)]{DOEIsotope2025}
{U.S. Department of Energy}.
\newblock Isotope {R\&D} and production {FY}~2025 congressional justification.
\newblock Technical report, DOE/SC, 2024.
\newblock URL
  \url{https://www.energy.gov/sites/default/files/2024-03/FY2025-PresidentsRequest-IRP.pdf}.

\bibitem[{American Nuclear Society}(2025)]{ANSIsotopeSupply}
{American Nuclear Society}.
\newblock Experts talk on developing the isotope supply chain.
\newblock \emph{Nuclear Newswire}, October 2025.

\bibitem[King(2024)]{ChemWorld2024}
A.~King.
\newblock The race to commercialise nuclear-powered batteries.
\newblock \emph{Chemistry World}, November 2024.
\newblock URL
  \url{https://www.chemistryworld.com/news/the-race-to-commercialise-nuclear-powered-batteries/4020505.article}.

\bibitem[Rutkowski et~al.(2025)Rutkowski, Harter, and Parisi]{Rutkowski2025}
A.~Rutkowski, J.~Harter, and J.~Parisi.
\newblock Scalable chrysopoeia via $(n,2n)$ reactions driven by high-flux
  fusion neutrons.
\newblock \emph{arXiv preprint arXiv:2507.13461}, 2025.

\bibitem[Gilbert et~al.(2024)]{Gilbert2024}
M.~R. Gilbert et~al.
\newblock Neutron activation and transmutation in fusion blankets.
\newblock \emph{Nucl. Fusion}, 64:\penalty0 036021, 2024.

\bibitem[Parisi et~al.(2025)Parisi, Rutkowski, Harter, Schwartz, and
  Chen]{Parisi2025}
J.~F. Parisi, A.~Rutkowski, J.~Harter, J.~A. Schwartz, and S.~Chen.
\newblock Production of high-specific-activity radioisotopes using high-energy
  fusion neutrons.
\newblock \emph{arXiv preprint}, 2025.
\newblock URL \url{https://arxiv.org/abs/2511.02814}.

\bibitem[Evitts et~al.(2025)Evitts, Miller, Pieve, Turner, and
  Borini]{Evitts2025}
L.~J. Evitts, P.~W. Miller, C.~Da Pieve, A.~Turner, and S.~Borini.
\newblock Theoretical novel medical isotope production with deuterium-tritium
  fusion technology.
\newblock \emph{Appl. Radiat. Isot.}, 226:\penalty0 112163, 2025.
\newblock URL \url{https://doi.org/10.1016/j.apradiso.2025.112163}.

\bibitem[Romano et~al.(2015)Romano, Horelik, Herman, Nelson, Forget, and
  Smith]{openmc}
P.~K. Romano, N.~E. Horelik, B.~R. Herman, A.~G. Nelson, B.~Forget, and
  K.~Smith.
\newblock {OpenMC}: a state-of-the-art {Monte Carlo} code for research and
  development.
\newblock \emph{Ann. Nucl. Energy}, 82:\penalty0 90--97, 2015.
\newblock URL \url{https://doi.org/10.1016/j.anucene.2014.07.048}.

\bibitem[Flicker et~al.(1964)Flicker, Loferski, and
  Elleman]{flicker1964construction}
H~Flicker, JJ~Loferski, and TS~Elleman.
\newblock Construction of a promethium-147 atomic battery.
\newblock \emph{IEEE Transactions on Electron Devices}, 11\penalty0
  (1):\penalty0 2--8, 1964.
\newblock URL \url{https://doi.org/10.1109/T-ED.1964.15271}.

\bibitem[Yakubova(2010)]{yakubova2010nuclear}
Galina~Nikolayevna Yakubova.
\newblock \emph{Nuclear batteries with tritium and promethium-147 radioactive
  sources}.
\newblock University of Illinois at Urbana-Champaign, 2010.

\bibitem[Kavetskiy et~al.(2011)Kavetskiy, Yakubova, Yousaf, Bower, Robertson,
  and Garnov]{kavetskiy2011efficiency}
A~Kavetskiy, G~Yakubova, SM~Yousaf, K~Bower, JD~Robertson, and A~Garnov.
\newblock Efficiency of pm-147 direct charge radioisotope battery.
\newblock \emph{Applied Radiation and Isotopes}, 69\penalty0 (5):\penalty0
  744--748, 2011.

\bibitem[Waris et~al.(2016)Waris, Kusumawati, Alfarobi, Aji, and
  Basar]{waris2016preliminary}
A~Waris, Y~Kusumawati, AS~Alfarobi, IK~Aji, and K~Basar.
\newblock Preliminary design of betavoltaic battery using co-60 and pm-147 with
  gaas substrate.
\newblock In \emph{AIP Conference Proceedings}, volume 1719, page 030053. AIP
  Publishing LLC, 2016.

\bibitem[Broderick et~al.(2019)Broderick, Lusk, Hinderer, Griswold, Boll,
  Garland, Heilbronn, and Mirzadeh]{broderick2019reactor}
Kathleen Broderick, Rita Lusk, James Hinderer, Justin Griswold, Rose Boll, Marc
  Garland, Lawrence Heilbronn, and Saed Mirzadeh.
\newblock Reactor production of promethium-147.
\newblock \emph{Applied Radiation and Isotopes}, 144:\penalty0 54--63, 2019.
\newblock URL \url{https://doi.org/10.1016/j.apradiso.2018.10.025}.

\bibitem[Jurewitsch et~al.(1998)Jurewitsch, Boody, Fortov, and
  Hoepfl]{Jurewitsch1998radionuclide}
Wladimir Jurewitsch, Frederick~P. Boody, Wladimir~E. Fortov, and Reinhard
  Hoepfl.
\newblock Super-compact radionuclide battery useful for spacecraft contains
  radionuclide dust particles suspended in a gas or plasma, 7 1998.
\newblock URL \url{https://patents.google.com/patent/DE19833648A1}.
\newblock Application number DE1998133648. Priority date 1997-09-01. Status:
  Withdrawn.

\bibitem[Xu et~al.(2020)Xu, Liu, and Tang]{xu2020radioluminescent}
Zhiheng Xu, Yunpeng Liu, and Xiaobin Tang.
\newblock Radioluminescent nuclear battery technology development for space
  exploration.
\newblock \emph{Advances in Astronautics Science and Technology}, 3\penalty0
  (2):\penalty0 125--131, 2020.
\newblock URL \url{https://doi.org/10.1007/s42423-020-00067-w}.

\bibitem[Tang et~al.(2012)Tang, Ding, Liu, and Chen]{tang2012optimization}
XiaoBin Tang, Ding Ding, YunPeng Liu, and Da~Chen.
\newblock Optimization design and analysis of si-63ni betavoltaic battery.
\newblock \emph{Science China Technological Sciences}, 55\penalty0
  (4):\penalty0 990--996, 2012.
\newblock URL \url{https://doi.org/10.1007/s11431-012-4752-6}.

\bibitem[Munson et~al.(2015)Munson, Arif, Streque, Belahsene, Martinez,
  Ramdane, El~Gmili, Salvestrini, Voss, and Ougazzaden]{munson2015model}
Charles~E Munson, Muhammad Arif, Jeremy Streque, Sofiane Belahsene, Anthony
  Martinez, Abderrahim Ramdane, Youssef El~Gmili, Jean-Paul Salvestrini, Paul~L
  Voss, and Abdallah Ougazzaden.
\newblock Model of ni-63 battery with realistic pin structure.
\newblock \emph{Journal of applied physics}, 118\penalty0 (10):\penalty0
  105101, 2015.
\newblock URL \url{https://doi.org/10.1063/1.4930870}.

\bibitem[Abdou et~al.(2021)]{Abdou2021}
M.~Abdou et~al.
\newblock Physics and technology considerations for the deuterium-tritium fuel
  cycle and conditions for tritium fuel self sufficiency.
\newblock \emph{Nucl. Fusion}, 61:\penalty0 013001, 2021.
\newblock URL \url{https://doi.org/10.1088/1741-4326/abbf35}.

\bibitem[P{\'a}lffy et~al.(2007)P{\'a}lffy, Evers, and
  Keitel]{palffy2007isomer}
Adriana P{\'a}lffy, J{\"o}rg Evers, and Christoph~H Keitel.
\newblock Isomer triggering via nuclear excitation by electron capture.
\newblock \emph{Physical review letters}, 99\penalty0 (17):\penalty0 172502,
  2007.

\bibitem[Arnquist et~al.(2023)Arnquist, Avignone~III, Barabash, Barton,
  Bhimani, Blalock, Bos, Busch, Buuck, Caldwell,
  et~al.]{arnquist2023constraints}
IJ~Arnquist, FT~Avignone~III, AS~Barabash, CJ~Barton, KH~Bhimani, E~Blalock,
  B~Bos, M~Busch, M~Buuck, TS~Caldwell, et~al.
\newblock Constraints on the decay of ta 180 m.
\newblock \emph{Physical Review Letters}, 131\penalty0 (15):\penalty0 152501,
  2023.

\bibitem[Ding et~al.(2026)Ding, Jia, Guo, Zhou, Wu, Zeng, Wang, Qiang, Xu,
  Fang, et~al.]{ding2026isomer}
B~Ding, CX~Jia, S~Guo, XH~Zhou, YB~Wu, FF~Zeng, JG~Wang, YH~Qiang, SW~Xu,
  YD~Fang, et~al.
\newblock Isomer depletion of mo 93 m triggered by inelastic nuclear scattering
  rather than nuclear excitation by electron capture.
\newblock \emph{Physical Review Letters}, 136\penalty0 (5):\penalty0 052502,
  2026.

\bibitem[Parisi and Schiller(2026)]{parisi2026neutronvalue}
J.~F. Parisi and K.~Schiller.
\newblock The value and cost of fusion neutrons.
\newblock \emph{arXiv preprint arXiv:2603.00835}, 2026.

\bibitem[Brown et~al.(2018)Brown, Chadwick, Capote, Kahler, Trkov, Herman,
  Sonzogni, Danon, Carlson, Dunn, Smith, Hale, Arbanas, Arcilla, Bates, Beck,
  Becker, Brown, Casperson, Conlin, Cullen, Descalle, Firestone, Gaines, Guber,
  Hawari, Holmes, Johnson, Kawano, Kiedrowski, Koning, Kopecky, Leal, Lestone,
  Lubitz, Damian, Mattoon, McCutchan, Mughabghab, Navratil, Neudecker, Nobre,
  Noguere, Paris, Pigni, Plompen, Pritychenko, Pronyaev, Roubtsov, Rochman,
  Romano, Schillebeeckx, Simakov, Sin, Sirakov, Sleaford, Sobes, Soukhovitskii,
  Stetcu, Talou, Thompson, van~der Marck, Welser-Sherrill, Wiarda, White,
  Wormald, Wright, Zerkle, \v{Z}erovnik, and Zhu]{Brown2018}
D.~A. Brown, M.~B. Chadwick, R.~Capote, A.~C. Kahler, A.~Trkov, M.~W. Herman,
  A.~A. Sonzogni, Y.~Danon, A.~D. Carlson, M.~Dunn, D.~L. Smith, G.~M. Hale,
  G.~Arbanas, R.~Arcilla, C.~R. Bates, B.~Beck, B.~Becker, F.~Brown, R.~J.
  Casperson, J.~Conlin, D.~E. Cullen, M.-A. Descalle, R.~Firestone, T.~Gaines,
  K.~H. Guber, A.~I. Hawari, J.~Holmes, T.~D. Johnson, T.~Kawano, B.~C.
  Kiedrowski, A.~J. Koning, S.~Kopecky, L.~Leal, J.~P. Lestone, C.~Lubitz,
  J.~I.~Marquez Damian, C.~M. Mattoon, E.~A. McCutchan, S.~Mughabghab,
  P.~Navratil, D.~Neudecker, G.~P.~A. Nobre, G.~Noguere, M.~Paris, M.~T. Pigni,
  A.~J. Plompen, B.~Pritychenko, V.~G. Pronyaev, D.~Roubtsov, D.~Rochman,
  P.~Romano, P.~Schillebeeckx, S.~Simakov, M.~Sin, I.~Sirakov, B.~Sleaford,
  V.~Sobes, E.~S. Soukhovitskii, I.~Stetcu, P.~Talou, I.~Thompson, S.~van~der
  Marck, L.~Welser-Sherrill, D.~Wiarda, M.~White, J.~L. Wormald, R.~Q. Wright,
  M.~Zerkle, G.~\v{Z}erovnik, and Y.~Zhu.
\newblock {ENDF/B-VIII.0}: The 8th major release of the nuclear reaction data
  library with {CIELO}-project cross sections, new standards and thermal
  scattering data.
\newblock \emph{Nucl. Data Sheets}, 148:\penalty0 1--142, 2018.
\newblock \doi{10.1016/j.nds.2018.02.001}.

\bibitem[{U.S. Nuclear Regulatory Commission}(2023)]{nrc_secy_2023_0001}
{U.S. Nuclear Regulatory Commission}.
\newblock Staff requirements -- {SECY-23-0001} -- options for advancing the
  {Commission's} strategy for the regulation of fusion energy systems.
\newblock {ADAMS} Accession No. {ML23103A099}, April 2023.
\newblock URL \url{https://www.nrc.gov/docs/ML2310/ML23103A099.pdf}.

\bibitem[{U.S. Nuclear Regulatory Commission}(2024{\natexlab{a}})]{cfr_10_30}
{U.S. Nuclear Regulatory Commission}.
\newblock Rules of general applicability to domestic licensing of byproduct
  material.
\newblock Title 10, Code of Federal Regulations, Part 30, 2024{\natexlab{a}}.
\newblock URL
  \url{https://www.nrc.gov/reading-rm/doc-collections/cfr/part030/}.

\bibitem[{U.S. Nuclear Regulatory Commission}(2024{\natexlab{b}})]{cfr_10_61}
{U.S. Nuclear Regulatory Commission}.
\newblock Licensing requirements for land disposal of radioactive waste.
\newblock Title 10, Code of Federal Regulations, Part 61, 2024{\natexlab{b}}.
\newblock URL
  \url{https://www.nrc.gov/reading-rm/doc-collections/cfr/part061/}.

\bibitem[{U.S. Nuclear Regulatory
  Commission}(2024{\natexlab{c}})]{cfr_10_61_55}
{U.S. Nuclear Regulatory Commission}.
\newblock Waste classification.
\newblock Title 10, Code of Federal Regulations, Part 61.55,
  2024{\natexlab{c}}.
\newblock URL
  \url{https://www.nrc.gov/reading-rm/doc-collections/cfr/part061/part061-0055.html}.

\bibitem[{U.S. Department of Energy}(2021)]{doe_order_435_1}
{U.S. Department of Energy}.
\newblock {DOE Order 435.1} -- radioactive waste management.
\newblock Chg 2, Office of Environmental Management, 2021.
\newblock URL
  \url{https://www.directives.doe.gov/directives-documents/400-series/0435.1-BOrder-chg2-PgChg}.

\bibitem[{U.S. Nuclear Regulatory Commission}(2024{\natexlab{d}})]{cfr_10_32}
{U.S. Nuclear Regulatory Commission}.
\newblock Specific domestic licenses to manufacture or transfer certain items
  containing byproduct material.
\newblock Title 10, Code of Federal Regulations, Part 32, 2024{\natexlab{d}}.
\newblock URL
  \url{https://www.nrc.gov/reading-rm/doc-collections/cfr/part032/}.

\bibitem[Boschi et~al.(2019)Boschi, Uccelli, and Martini]{boschi2019picture}
Alessandra Boschi, Licia Uccelli, and Petra Martini.
\newblock A picture of modern tc-99m radiopharmaceuticals: Production,
  chemistry, and applications in molecular imaging.
\newblock \emph{Applied Sciences}, 9\penalty0 (12):\penalty0 2526, 2019.

\end{thebibliography}

\appendix

\section{Depletion simulation with Pm + Sm extraction}
\label{app:tokamak_PmSm}

\begin{figure}[b!]
\centering
\includegraphics[width=\columnwidth]{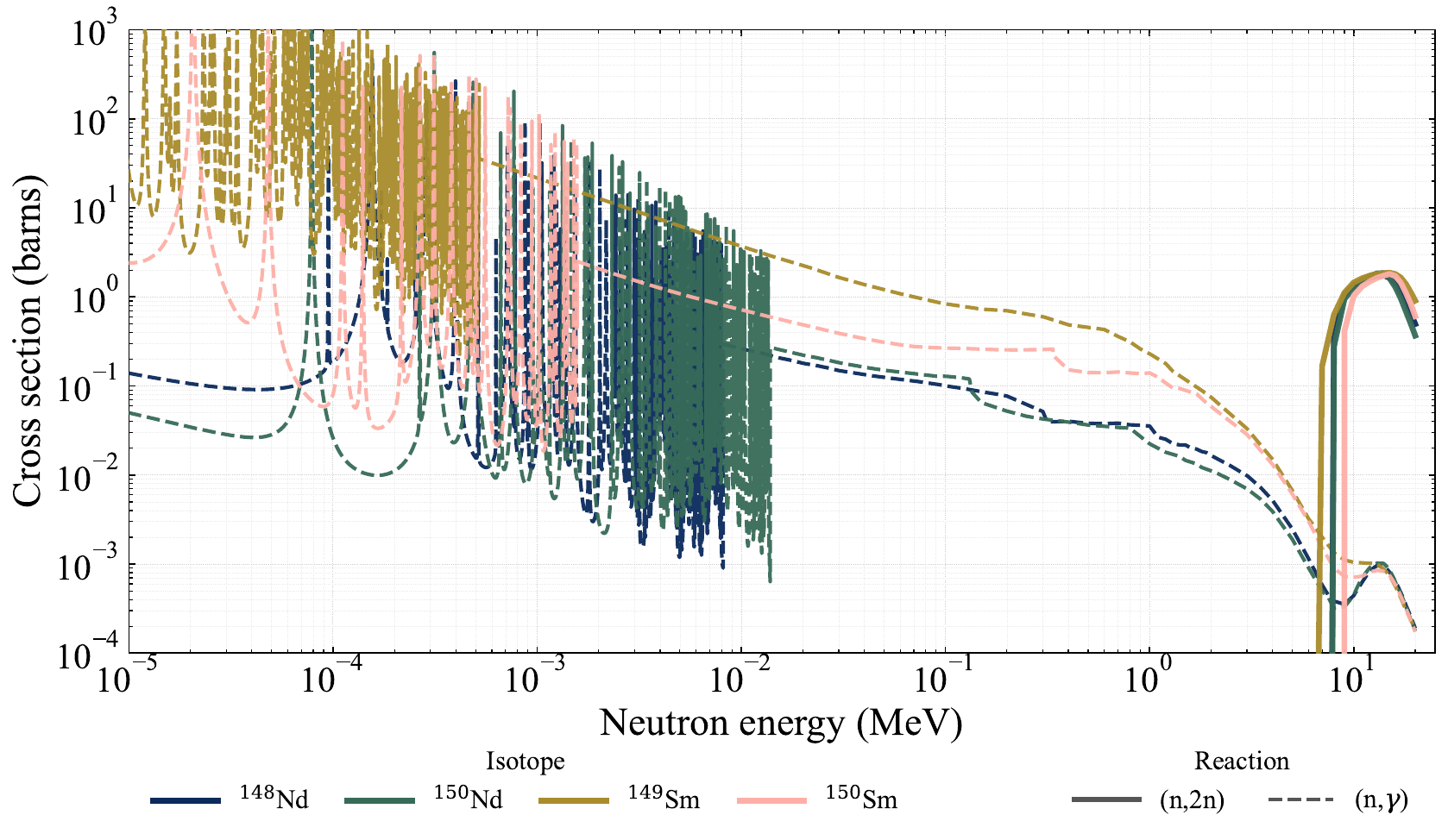}
\caption{(n,$\gamma$) and (n,2n) cross sections on some Nd and Sm isotopes.}
\label{fig:ng_XS}
\end{figure}

\begin{figure*}[tb!]
\centering
\includegraphics[width=\textwidth]{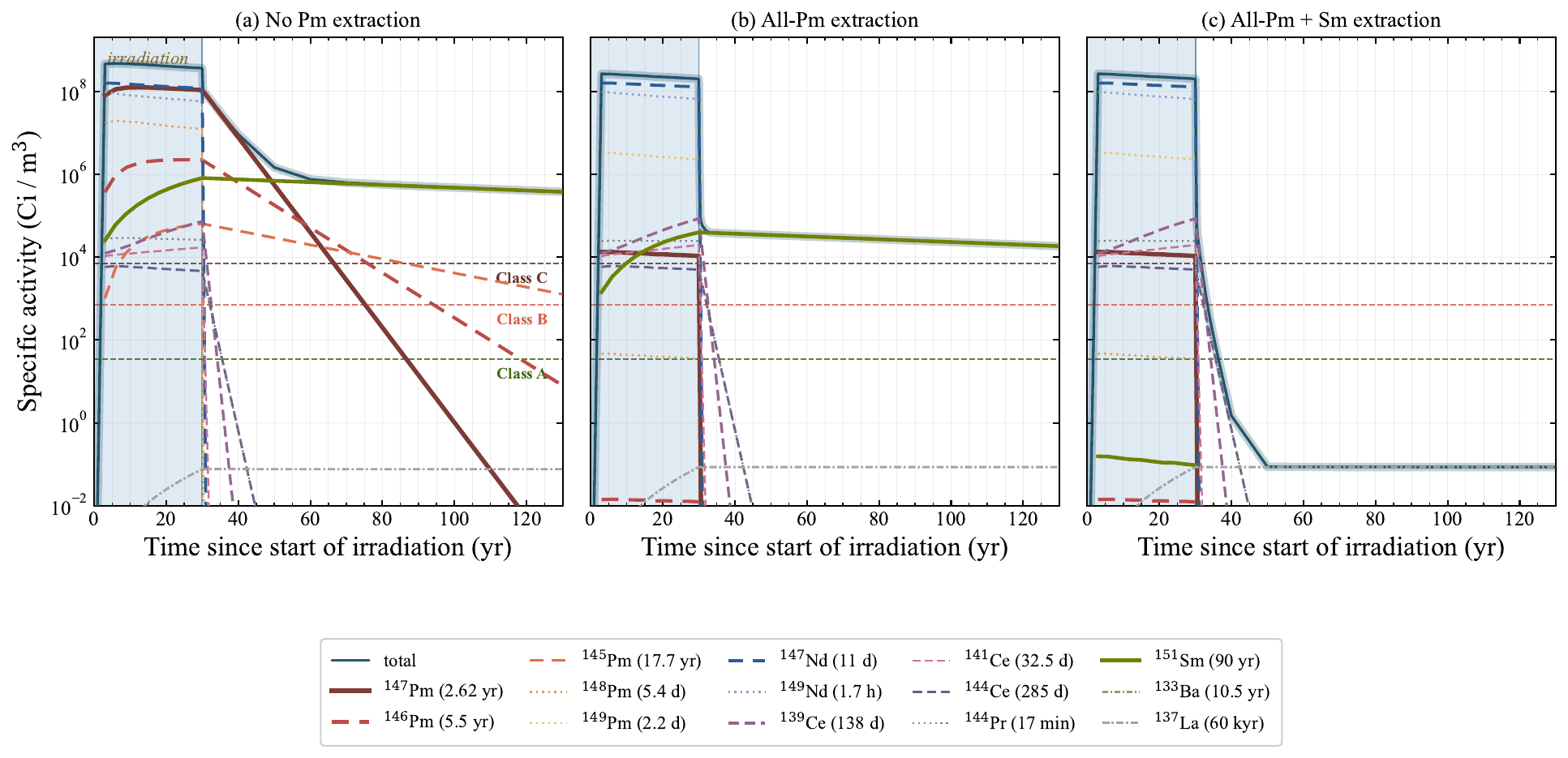}
\caption{Specific activity in the enriched ${}^{148}$Nd blanket as a function of time since start of irradiation, for three extraction schemes. Shaded blue region is the 30 year irradiation window. Horizontal dashed lines are the generic $\beta/\gamma$ Class A, B, and C limits for activated ${}^{63}$Ni metal waste. The no-extraction and all-Pm cases remain above Class C (GTCC) for centuries; Pm + Sm dual extraction drops the blanket below Class C within a year of shutdown. Data: ENDF/B-VIII.0~\cite{Brown2018}}
\label{fig:waste_blanket}
\end{figure*}

The OpenMC depletion simulation described in this appendix removes both Pm and Sm continuously from the blanket during irradiation. This continuous removal serves to minimize burnup of $^{147}$Pm and to minimize long-term activation products in the blanket. The OpenMC simulation uses the large-tokamak DAGMC CAD geometry of~\cite{Rutkowski2025} (1.5 GW fusion power, major radius $R = 4.2$~m, minor radius $a = 1.2$~m, elongation $\kappa = 1.6$, triangularity $\delta = 0.25$, Shafranov shift $0.45$). We use 90\%at enriched $^{148}$Nd, and the non-$^{148}$Nd isotopic abundances are rescaled linearly according to their natural abundances. Here, we focus solely on the enriched $^{148}$Nd blanket layer (without Li) shown in \Cref{fig:simulation_setup}; we do not analyze the FLiBe layer.

Sm extraction is required, otherwise it would leave a large quantity of ${}^{151}$Sm ($t_\mathrm{1/2}=$95 years) in the blanket at shutdown, produced by chains such as
\begin{equation}
\begin{aligned}
& {}^{148}\mathrm{Nd}(\mathrm{n},\gamma){}^{149}\mathrm{Nd} \xrightarrow{\beta^-} {}^{149}\mathrm{Pm} \\
& \xrightarrow{\beta^-} {}^{149}\mathrm{Sm}(\mathrm{n},\gamma){}^{150}\mathrm{Sm}(\mathrm{n},\gamma){}^{151}\mathrm{Sm}.
\end{aligned}
\label{eq:sm_backdoor}
\end{equation}
and
\begin{equation}
\begin{aligned}
& {}^{148}\mathrm{Nd}(\mathrm{n},2 \mathrm{n}){}^{149}\mathrm{Nd} \xrightarrow{\beta^-} {}^{149}\mathrm{Pm} \\
& \xrightarrow{\beta^-} {}^{149}\mathrm{Sm}(\mathrm{n},\gamma){}^{150}\mathrm{Sm}(\mathrm{n},\gamma){}^{151}\mathrm{Sm}.
\end{aligned}
\label{eq:sm_backdoor2}
\end{equation}
We show some (n,$\gamma$) and (n,2n) cross sections in \Cref{fig:ng_XS}. Notably, the Sm(n,$\gamma$) cross sections are very high, exceeding 100 millibarn for neutron energy lower than $\sim$2 MeV. This is why Sm removal is so effective at reducing accumulation of ${}^{151}\mathrm{Sm}$ in the blanket. Extraction of Sm interrupts this chain before ${}^{149}$Sm or ${}^{150}$Sm atom can neutron capture, reducing the in-blanket ${}^{151}$Sm inventory by five orders of magnitude compared with leaving Sm in the blanket.

\begin{table*}[t!]
\centering
\caption{Complete downstream nuclide inventory of a ${}^{148}$Nd channel (100~mm thick, 90\% enriched) at the end of 30~yr of irradiation with continuous extraction of both Pm and Sm isotopes. The total fusion power is 1.5 GW. Bolded entries: desired product (${}^{147}$Pm), long-lived contaminants (${}^{146,145}$Pm, ${}^{137}$La), and residual ${}^{151}$Sm.}
\label{tab:tokamak_inventory_PmSm}
\scriptsize
\setlength{\tabcolsep}{5pt}
\begin{tabular}{@{}lrrrrr@{}}
\toprule
Nuclide & $\thalf$ & Atoms & Mass & Activity (Ci) & Decay heat (W) \\
\midrule
$^{148}$Nd & stable & $4.97{\cdot}10^{29}$ & 122\,t & --- & --- \\
$^{142}$Nd & stable & $1.90{\cdot}10^{28}$ & 4.48\,t & --- & --- \\
$^{144}$Nd & stable & $1.78{\cdot}10^{28}$ & 4.25\,t & $4.61{\cdot}10^{-6}$ & --- \\
$^{146}$Nd & stable & $1.48{\cdot}10^{28}$ & 3.59\,t & --- & --- \\
$^{143}$Nd & stable & $9.18{\cdot}10^{27}$ & 2.18\,t & --- & --- \\
$^{145}$Nd & stable & $6.60{\cdot}10^{27}$ & 1.59\,t & --- & --- \\
$^{141}$Pr & stable & $3.42{\cdot}10^{27}$ & 800\,kg & --- & --- \\
$^{150}$Nd & stable & $2.77{\cdot}10^{27}$ & 691\,kg & --- & --- \\
$^{140}$Ce & stable & $3.16{\cdot}10^{26}$ & 73.5\,kg & --- & --- \\
$^{147}$Nd & 11\,d & $1.79{\cdot}10^{26}$ & 43.7\,kg & $3.53{\cdot}10^{9}$ & $8.83{\cdot}10^{6}$ \\
$^{139}$La & stable & $3.30{\cdot}10^{25}$ & 7.61\,kg & --- & --- \\
$^{142}$Ce & stable & $2.63{\cdot}10^{24}$ & 620\,g & --- & --- \\
$^{138}$La & stable & $2.34{\cdot}10^{24}$ & 536\,g & $1.35{\cdot}10^{-5}$ & --- \\
$^{139}$Ce & 138\,d & $1.50{\cdot}10^{24}$ & 346\,g & $2.36{\cdot}10^{6}$ & $2.60{\cdot}10^{3}$ \\
\textbf{$^{147}$Pm} & \textbf{2.6\,yr} & \textbf{$1.31{\cdot}10^{24}$} & \textbf{319\,g} & \textbf{$2.96{\cdot}10^{5}$} & \textbf{$1.09{\cdot}10^{2}$} \\
$^{149}$Pm & 2.2\,d & $6.38{\cdot}10^{23}$ & 158\,g & $6.26{\cdot}10^{7}$ & $1.40{\cdot}10^{5}$ \\
$^{149}$Nd & 1.7\,h & $5.94{\cdot}10^{23}$ & 147\,g & $1.79{\cdot}10^{9}$ & $1.02{\cdot}10^{7}$ \\
$^{138}$Ce & stable & $5.82{\cdot}10^{23}$ & 133\,g & --- & --- \\
\textbf{$^{137}$La} & 60000\,yr & $2.41{\cdot}10^{23}$ & \textbf{54.8\,g} & $2.38{\cdot}10^{0}$ & $4.34{\cdot}10^{-4}$ \\
$^{144}$Ce & 285\,d & $1.82{\cdot}10^{23}$ & 43.6\,g & $1.39{\cdot}10^{5}$ & $9.88{\cdot}10^{1}$ \\
$^{141}$Ce & 33\,d & $8.14{\cdot}10^{22}$ & 19.1\,g & $5.43{\cdot}10^{5}$ & $8.73{\cdot}10^{2}$ \\
$^{142}$Pr & 19\,h & $5.95{\cdot}10^{22}$ & 14\,g & $1.62{\cdot}10^{7}$ & $8.33{\cdot}10^{4}$ \\
$^{141}$Nd & 2.5\,h & $5.45{\cdot}10^{22}$ & 12.8\,g & $1.14{\cdot}10^{8}$ & $6.15{\cdot}10^{4}$ \\
$^{137}$Ba & stable & $5.14{\cdot}10^{22}$ & 11.7\,g & --- & --- \\
$^{143}$Pr & 14\,d & $3.36{\cdot}10^{22}$ & 7.98\,g & $5.37{\cdot}10^{5}$ & $1.00{\cdot}10^{3}$ \\
$^{149}$Sm & stable & $2.32{\cdot}10^{22}$ & 5.73\,g & --- & --- \\
$^{136}$Ba & stable & $9.47{\cdot}10^{21}$ & 2.14\,g & --- & --- \\
$^{151}$Pm & 1.2\,d & $5.90{\cdot}10^{21}$ & 1.48\,g & $1.08{\cdot}10^{6}$ & $4.19{\cdot}10^{3}$ \\
$^{145}$Pr & 6\,h & $3.61{\cdot}10^{21}$ & 870\,mg & $3.14{\cdot}10^{6}$ & $1.29{\cdot}10^{4}$ \\
$^{138}$Ba & stable & $3.04{\cdot}10^{21}$ & 697\,mg & --- & --- \\
$^{143}$Ce & 1.4\,d & $1.14{\cdot}10^{21}$ & 271\,mg & $1.80{\cdot}10^{5}$ & $7.80{\cdot}10^{2}$ \\
$^{151}$Nd & 12\,min & $6.78{\cdot}10^{20}$ & 170\,mg & $1.70{\cdot}10^{7}$ & $1.54{\cdot}10^{5}$ \\
$^{135}$Ba & stable & $6.09{\cdot}10^{20}$ & 136\,mg & --- & --- \\
$^{140}$La & 1.7\,d & $4.37{\cdot}10^{20}$ & 102\,mg & $5.64{\cdot}10^{4}$ & $9.53{\cdot}10^{2}$ \\
\textbf{$^{151}$Sm} & \textbf{90\,yr} & \textbf{$4.00{\cdot}10^{20}$} & \textbf{100\,mg} & \textbf{$2.64{\cdot}10^{0}$} & \textbf{$3.10{\cdot}10^{-4}$} \\
$^{140}$Pr & 3.4\,min & $1.98{\cdot}10^{20}$ & 46\,mg & $1.82{\cdot}10^{7}$ & $1.18{\cdot}10^{5}$ \\
$^{147}$Sm & stable & $1.09{\cdot}10^{20}$ & 26.7\,mg & --- & --- \\
$^{144}$Pr & 17\,min & $3.75{\cdot}10^{19}$ & 8.97\,mg & $6.78{\cdot}10^{5}$ & $4.97{\cdot}10^{3}$ \\
$^{134}$Ba & stable & $3.59{\cdot}10^{19}$ & 7.99\,mg & --- & --- \\
$^{145}$Ce & 3\,min & $2.90{\cdot}10^{19}$ & 6.98\,mg & $3.01{\cdot}10^{6}$ & $2.76{\cdot}10^{4}$ \\
$^{148}$Pm & 5.4\,d & $2.32{\cdot}10^{19}$ & 5.7\,mg & $9.36{\cdot}10^{2}$ & $7.25{\cdot}10^{0}$ \\
$^{151}$Eu & stable & $2.00{\cdot}10^{19}$ & 5.02\,mg & --- & --- \\
$^{148}$Pr & 2.3\,min & $1.77{\cdot}10^{19}$ & 4.34\,mg & $2.41{\cdot}10^{6}$ & $4.73{\cdot}10^{4}$ \\
$^{146}$Pr & 24\,min & $1.28{\cdot}10^{19}$ & 3.11\,mg & $1.66{\cdot}10^{5}$ & $2.32{\cdot}10^{3}$ \\
$^{150}$Pm & 2.7\,h & $5.92{\cdot}10^{18}$ & 1.47\,mg & $1.14{\cdot}10^{4}$ & $1.51{\cdot}10^{2}$ \\
$^{150}$Sm & stable & $4.90{\cdot}10^{18}$ & 1.22\,mg & --- & --- \\
$^{137}$Ce & 9\,h & $5.34{\cdot}10^{18}$ & 1.21\,mg & $3.09{\cdot}10^{3}$ & $8.20{\cdot}10^{-1}$ \\
\textbf{$^{146}$Pm} & \textbf{5.5\,yr} & \textbf{$3.19{\cdot}10^{18}$} & \textbf{774\,$\mu$g} & \textbf{$3.43{\cdot}10^{-1}$} & \textbf{$1.71{\cdot}10^{-3}$} \\
$^{133}$Ba & 11\,yr & $1.06{\cdot}10^{18}$ & 234\,$\mu$g & $5.97{\cdot}10^{-2}$ & $1.51{\cdot}10^{-4}$ \\
$^{147}$Pr & 13\,min & $4.83{\cdot}10^{17}$ & 118\,$\mu$g & $1.13{\cdot}10^{4}$ & $1.15{\cdot}10^{2}$ \\
$^{148}$Sm & stable & $4.12{\cdot}10^{17}$ & 101\,$\mu$g & --- & --- \\
$^{139}$Ba & 1.4\,h & $1.13{\cdot}10^{17}$ & 26.1\,$\mu$g & $4.26{\cdot}10^{2}$ & $2.41{\cdot}10^{0}$ \\
$^{152}$Sm & stable & $1.02{\cdot}10^{17}$ & 25.7\,$\mu$g & --- & --- \\
$^{132}$Ba & stable & $3.93{\cdot}10^{16}$ & 8.62\,$\mu$g & --- & --- \\
$^{144\mathrm{m}}$Pr & 7.2\,min & $3.13{\cdot}10^{16}$ & 7.48\,$\mu$g & $1.36{\cdot}10^{3}$ & $4.47{\cdot}10^{-1}$ \\
$^{147}$Ce & 0.94\,min & $2.28{\cdot}10^{16}$ & 5.56\,$\mu$g & $7.57{\cdot}10^{3}$ & $9.91{\cdot}10^{1}$ \\
$^{140}$Ba & 13\,d & $1.11{\cdot}10^{16}$ & 2.58\,$\mu$g & $1.88{\cdot}10^{-1}$ & $5.76{\cdot}10^{-4}$ \\
$^{142}$La & 1.5\,h & $1.02{\cdot}10^{16}$ & 2.4\,$\mu$g & $3.49{\cdot}10^{1}$ & $6.39{\cdot}10^{-1}$ \\
$^{146}$Ce & 13\,min & $7.10{\cdot}10^{15}$ & 1.72\,$\mu$g & $1.64{\cdot}10^{2}$ & $5.93{\cdot}10^{-1}$ \\
$^{141}$La & 3.9\,h & $4.26{\cdot}10^{15}$ & 997\,ng & $5.66{\cdot}10^{0}$ & $3.40{\cdot}10^{-2}$ \\
$^{152}$Pm & 4.1\,min & $3.06{\cdot}10^{15}$ & 772\,ng & $2.32{\cdot}10^{2}$ & $2.32{\cdot}10^{0}$ \\
$^{150}$Pr & 0.1\,min & $1.79{\cdot}10^{15}$ & 446\,ng & $5.42{\cdot}10^{3}$ & $7.90{\cdot}10^{1}$ \\
$^{146}$Sm & stable & $3.28{\cdot}10^{14}$ & 79.6\,ng & --- & --- \\
\textbf{$^{145}$Pm} & \textbf{18\,yr} & \textbf{$1.93{\cdot}10^{14}$} & \textbf{46.5\,ng} & \textbf{$6.48{\cdot}10^{-6}$} & \textbf{---} \\
$^{141}$Ba & 18\,min & $3.08{\cdot}10^{13}$ & 7.21\,ng & $5.26{\cdot}10^{-1}$ & $5.85{\cdot}10^{-3}$ \\
$^{131}$Ba & 11\,d & $1.22{\cdot}10^{13}$ & 2.66\,ng & $2.30{\cdot}10^{-4}$ & --- \\
$^{149}$Pr & 2.3\,min & $1.07{\cdot}10^{13}$ & 2.65\,ng & $1.48{\cdot}10^{0}$ & $1.94{\cdot}10^{-2}$ \\
\bottomrule
\end{tabular}
\end{table*}

\begin{table*}[t!]
\centering
\caption{Cumulative atoms extracted from the channel over 30~yr of dual Pm + Sm online extraction. Masses are tallied at the moment of extraction; the short-lived Pm isotopes ($\thalf < 1$\,yr) $\beta^-$-decay to the corresponding Sm in the processing loop before the 30 year irradiation ends, and the downstream Pm/Sm chemical separation removes that Sm before the ${}^{147}$Pm product is sold. Pm isotopes form the target product (${}^{147}$Pm after chemical Pm/Sm separation); Sm isotopes are routed to a small concentrated Sm waste stream containing the bulk of the ${}^{151}$Sm that would otherwise accumulate in the Nd blanket via Sm(n,$\gamma$) chains.}
\label{tab:tokamak_PmSm_stream}
\small
\setlength{\tabcolsep}{5pt}
\begin{tabular}{@{}lcccc@{}}
\toprule
Stream & Nuclide & $\thalf$ & Cumulative mass extracted & Note \\
\midrule
Pm product & \textbf{$^{147}$Pm} & 2.6\,yr & 32.7\,t & \textbf{target battery fuel} \\
 & $^{149}$Pm & 2.2\,d & 17.9\,t & decays to ${}^{149}$Sm in loop \\
 & $^{151}$Pm & 1.2\,d & 176\,kg & decays to ${}^{151}$Sm in loop \\
 & $^{146}$Pm & 5.5\,yr & 75.7\,g & long-lived contaminant (via Pm-147(n,2n)) \\
 & $^{148}$Pm & 5.4\,d & 624\,g & decays to ${}^{148}$Sm in loop \\
 & $^{150}$Pm & 2.7\,h & 179\,g & decays to ${}^{150}$Sm in loop \\
 & $^{152}$Pm & 4.1\,min & 98.3\,mg & decays to ${}^{152}$Sm in loop \\
 & $^{145}$Pm & 18\,yr & 4.45\,mg & trace long-lived contaminant \\
\addlinespace[3pt]
Sm waste & $^{149}$Sm & stable & 648\,kg & stable \\
 & \textbf{$^{151}$Sm} & 90\,yr & 11.9\,kg & \textbf{waste/secondary battery fuel} \\
 & $^{147}$Sm & stable & 2.74\,kg & ${\sim}\,$stable (from Pm-147 decay) \\
 & $^{150}$Sm & stable & 148\,g & stable \\
 & $^{148}$Sm & stable & 11.1\,g & $\sim$stable \\
 & $^{146}$Sm & stable & 7.78\,mg & ${\sim}\,$stable \\
 & $^{152}$Sm & stable & 3.27\,g & stable \\
\bottomrule
\end{tabular}
\end{table*}

We next determine the waste disposal requirements for materials producing in a Nd blanket.

Radioactive inventories in a fusion blanket fall into two broad categories. The first is bulk activated material in the blanket channel, cubic-meter-scale volumes whose ultimate disposal pathway is not yet settled for commercial fusion. The 2023 NRC memo SECY-23-0001~\cite{nrc_secy_2023_0001} directs that commercial fusion machines be regulated under 10 CFR Part 30 (as byproduct material)~\cite{cfr_10_30}, making it possible that activated structural waste will be routed into low-level waste (LLW), currently governed by 10 CFR 61~\cite{cfr_10_61} with its Class A/B/C concentration limits~\cite{cfr_10_61_55}. However final waste-acceptance criteria for fusion-specific activation products have not been codified and none of the dominant long-lived contributors here (${}^{151}$Sm, ${}^{146}$Pm, ${}^{145}$Pm) are tabulated in 10 CFR 61.55; it is likely that 10 CFR 61.55 will need to be expanded to include fusion-energy-relevant radioisotopes. State-level disposal sites such as Texas and Utah also impose their own acceptance criteria, and DOE-funded fusion facilities operate under a parallel framework (DOE Order 435.1~\cite{doe_order_435_1}) rather than 10 CFR 61. We adopt the 10 CFR 61.55 limits with a ``${}^{63}$Ni in activated metal'' analogue as a benchmark for comparison with current LLW practice, recognizing that the actual regulatory pathway may be refined before commercial fusion blanket facilities are deployed. 

LLW classification in the United States is governed by 10 CFR 61.55~\cite{cfr_10_61_55}. The (a)(7) sub-paragraph prescribes a ``sum-of-fractions'' test for mixtures of radionuclides. For waste containing radionuclides at specific activities $C_i$ (Ci m$^{-3}$), the quantity
\begin{equation}
\mathrm{SF}_X \;=\; \sum_i \frac{C_i}{L_{X,i}}
\label{eq:sof}
\end{equation}
is formed for each class $X \in \{A, B, C\}$, where $L_{X,i}$ is the class-$X$ concentration limit for nuclide $i$. The waste is Class $X$ if $\mathrm{SF}_X < 1$; if $\mathrm{SF}_C > 1$ the waste is greater than Class C (GTCC) and is not accepted for near-surface disposal in the US.

\Cref{fig:waste_blanket} shows the blanket specific activity during and after irradiation against these limits. We show depletion simulations for three cases: (a) no extraction, (b) Pm extraction, and (c) Pm and Sm extraction. Only the simultaneous extraction of Pm and Sm allows for blanket inventory that quickly falls below Class A, B, and C classification after irradiation.

The second category is sealed-source processing material, the Pm product stream (${\sim}\,4.5$\,m$^3$, containing the ${}^{147}$Pm battery fuel itself) and the Sm waste stream (${\sim}0.5$-$2$\,L of concentrated Sm metal after Pm/Sm chemical separation). Under current US practice these are handled under NRC byproduct-material rules (10 CFR 30~\cite{cfr_10_30}, with 10 CFR 32~\cite{cfr_10_32} governing manufacture and distribution of byproduct material in sealed sources) rather than as volumetric LLW. The regulatory controls are on total activity, shielding, containment, and transport packaging, not on a per cubic meter concentration limit. High specific activity is normal for such streams. Medical isotope systems such as ${}^{99}$Mo/${}^{99\mathrm{m}}$Tc generators~\cite{boschi2019picture} operate in this regime.

Of the long-lived nuclides in our inventory tables (${}^{151}$Sm, ${}^{146}$Pm, ${}^{145}$Pm, ${}^{137}$La, ${}^{133}$Ba), none is explicitly listed in 10 CFR 61.55. As an approximate benchmark for pure low-energy $\beta^-$ emitters embedded in a metallic matrix, we adopt the ``${}^{63}$Ni in activated metal'' entry from Table 1, with limits $L_A = 35~\text{Ci\,m}^{-3}$, $L_B = 700~\text{Ci\,m}^{-3}$, $L_C = 7{,}000~\text{Ci\,m}^{-3}$.

${}^{63}$Ni ($t_{1/2} = 100$\,yr, pure $\beta^-$ with 67 keV) is an especially good surrogate for ${}^{151}$Sm ($t_{1/2} = 90$\,yr, pure $\beta^-$ with 77 keV); it is a worse surrogate for ${}^{146}$Pm (mixed EC/$\beta^-$, $\gamma$ at 454 and 736 keV) and ${}^{145}$Pm (predominantly EC, $\gamma$ at $\sim 67$ and 72 keV). We apply the same ${}^{63}$Ni-analogue limits to all long-lived species in this analysis as a uniform benchmark, with the understanding that a formal licensee submission would replace them with nuclide-specific radiotoxicity weightings and that the regulator retains discretion over the final classification.

Applied to the Pm + Sm simulation at 100 yr post-shutdown, the long-lived specific activity of the channel ($\sum_i C_i \approx 8.6 \cdot 10^{-2}$\,Ci\,m$^{-3}$) is essentially entirely from ${}^{137}$La (built up by (n,2n) on trace ${}^{138}$La in the Nd feedstock), with the in-blanket ${}^{151}$Sm reduced to a negligible level by Sm extraction. ${}^{137}$La activity dominates $\sim12$ years after end of irradiation. The sum of fractions test gives $\mathrm{SF}_A = 2.5 \cdot 10^{-3}$, several hundred times below the Class A threshold. The spent blanket channel is Class A LLW from $\sim$6 years after shutdown.

The channel inventory \Cref{tab:tokamak_inventory_PmSm} and the Pm+Sm-stream table \Cref{tab:tokamak_PmSm_stream} come from a single coupled neutron-transport plus depletion simulation performed in OpenMC~v0.15.2~\cite{openmc}, with continuous online removal of both Pm and Sm isotopes from the blanket channel via transfer rates applied to the depletion simulation.

\section{Cross Sections} \label{app:more_cross_sections}

\begin{figure*}[t!]
\centering
\includegraphics[width=2\columnwidth]{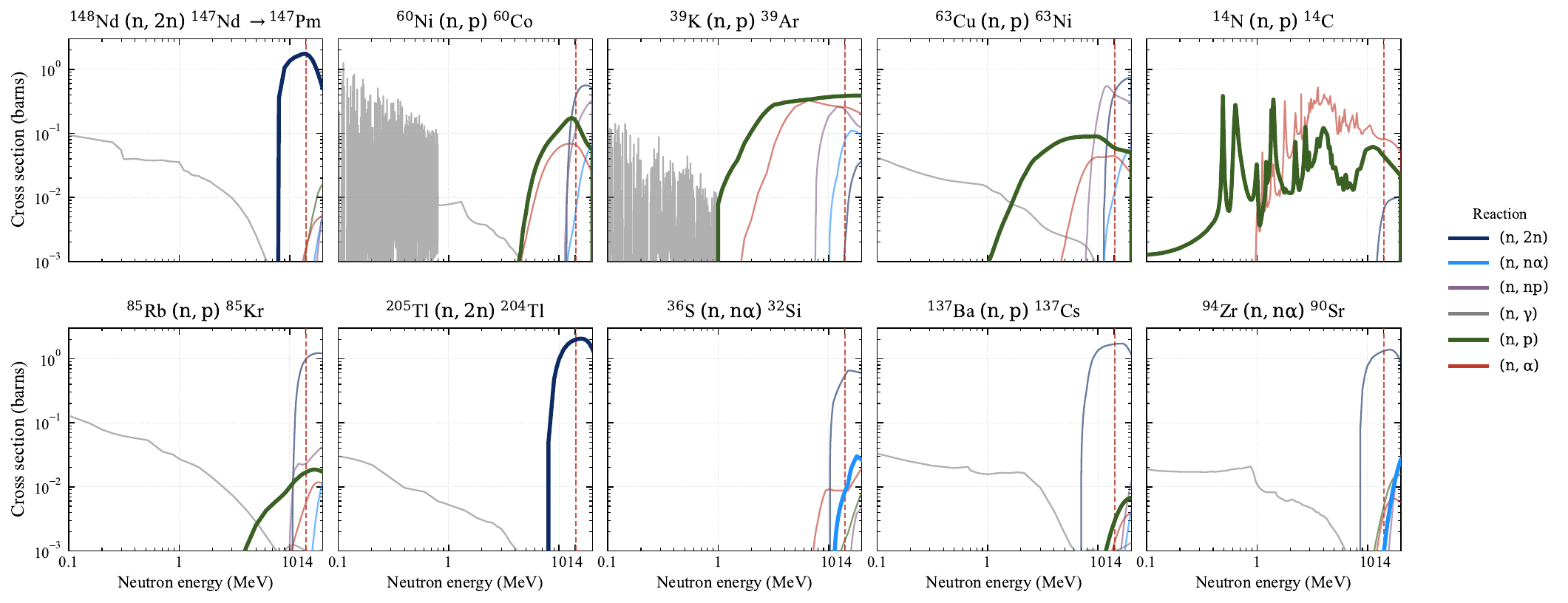}
\caption{Neutron-driven reaction cross sections for the dominant production pathways in \Cref{tab:battery}. Each panel highlights the labelled reaction (bold curve, colored by reaction category in the legend) on top of the other production-relevant channels on the same target (faint background curves). The red dashed vertical line marks the 14.1 D-T fusion neutron energy. Data: ENDF/B-VIII.0~\cite{Brown2018}.}
\label{fig:cross_sections}
\end{figure*}

In \Cref{fig:cross_sections} we show cross sections corresponding to radioisotope transmutation pathways with the highest production rates.

\end{document}